\documentclass[]{aastex631}

\usepackage{multirow}

\usepackage{amsmath}	

\begin{document}

\title{Forcing Planets to Evolve: Interactions Between Uranus and Neptune at Late Stages of Dynamical Evolution}

\author{Arcelia Hermosillo Ruiz}
\affiliation{Astronomy and Astrophysics, University of California Santa Cruz}
\author{Ruth Murray-Clay}
\affiliation{Astronomy and Astrophysics, University of California Santa Cruz}
\author{Kathryn Volk}
\affiliation{Planetary Science Institute}
\affiliation{Lunar and Planetary Laboratory, University of Arizona}
\author{Rosemary E. Pike}
\affiliation{Center for Astrophysics $|$ Harvard \& Smithsonian}



\begin{abstract}

In early Solar System numerical simulations, where chaos is a primary driver, it is difficult to explore parameter space in a systematic way. In such simulations, stable configurations are hard to come by, and often require special fine-tuning. In addition, it is infeasible to run suites of well-resolved, realistic simulations with a disk of massive particles to drive planetary evolution where enough particles remain to represent the transneptunian populations to robustly statistically compare with observations. To complement state of the art full N-body simulations, we develop a method to artificially control each planet's orbital elements independently from each other, which when carefully applied, can be used to test a wider suite of models. We modify two widely used publicly available N-body integrators: (1) the C code, \texttt{REBOUND} and (2) the FORTRAN code, \texttt{Mercury6.2}. We show how the application of specific fictitious forces within numerical integrators can be used to tightly control planetary evolution to more easily explore migration and orbital excitation and damping. This tool allows us to replicate the impact a massive planetesimal disk would have on the planets, without actually including the massive planetesimals, thus decreasing the chaos and simulation runtime. 
{We demonstrate the utility of this tool by applying it to the coupled orbital evolution of Uranus and Neptune, and show that Neptune's eccentricity damping and radial outward migration have the appropriate affect on Uranus' eccentricity.}

\end{abstract}



\section{Introduction}
Planets' orbits evolve through long and short term interactions with gas, planetesimals, and nearby planets \citep[e.g.,][]{Fernandez1984,Lee2002,Goldreich2004,chambers2009,Nesvorny:2018,morbidelli2020}.
When modeling planetary evolution caused by gravitational interactions with millions of small bodies, simplifications are necessary to address an otherwise computationally infeasible problem. 
A common approach in simplifying and decreasing the computational cost of an N-body simulation is to artificially evolve a planet's semimajor axis, eccentricity, and/or inclination. 
This is particularly important for science cases in which it is necessary to track the evolution of a statistically significant sample of planetesimals distinct from those driving the planetary evolution.
For example, when studying the evolution of the outer solar system, artificially controlling the giant planets evolution is helpful so that the particles representing the Trans Neptunian Objects (TNOs) can be massless, increasing the number that can be modeled and tracked, and their final dynamical configuration can be compared to currently observed distributions and used to constrain evolution models \citep[e.g][]{Hahn:2005,Levison2008,dawson2012,Nesvorny:2015,Nesvorny2016}.

Artificial evolution has been widely done with ``user-defined forces" where the code user defines an expression for the acceleration that will be added on top of the direct gravitational acceleration of the object due to other bodies in the simulation at each timestep \citep[e.g.,][]{malhotra1995,Hahn:2005,Wolff2012,Lee2002,AliDib2021}{}{}. In celestial mechanics, {the terms} ``force" and ``acceleration" are used interchangeably due to their proportional relationship for a given mass; for clarity, we will only refer to these as accelerations. {Additional accelerations are an existing functionality of widely used N-body integrators. The publicly available N-body codes \texttt{REBOUND} \citep[][]{Rein2012,rein2015,Rein2019}{}{} and \texttt{Mercury6.2} \citep[][]{Chambers:1999}{}{}, have the capacity to add an arbitrary acceleration to a particle of choice at every time-step, which operationally updates the particle's velocity every time-step. This allows a user to model physical forces on the planet and make it move in a prescribed way. A number of useful physical accelerations such as radiation forces, gas dynamical friction, and more, are already incorporated onto \texttt{REBOUND} through the library \texttt{REBOUNDx} {\citep[][]{Tamayo2020}}. The library also contains eccentricity and inclination damping, but the only free parameter the user can choose is a damping timescale. This combination of capabilities is enough to model physical forces on a particle. Neither code currently has the functionality to fully and independently evolve {any combination of} a planet's orbital elements; such an evolution requires both a ``user-defined velocity" and a ``user-defined acceleration."  User-defined velocities are not typically included in these codes because it is non-physical for the orbital elements to change {independently} from each other; {However,} they have been employed by \citet{Lee2002} to {independently} evolve semi-major axis and eccentricity and by \citet{Wolff2012} to independently evolve semi-major axis, eccentricity, and inclination.  }

In this paper we {present modifications to \texttt{REBOUND} and \texttt{Mercury6.2} that allow a user to evolve a particle's orbital elements independently from each other because of user-defined accelerations \textit{and} user-defined velocities. 
Our approach complements existing codes by allowing  arbitrary functions to be chosen for how the elements should evolve. 
While such evolution {may be} nonphysical, these modifications allow us to probe a large amount of parameter space and number of dynamical histories motivated by more realistic models to learn how {these evolutions} impact the structure and dynamics of systems.}

{The procedure described in this paper is most useful when
\begin{enumerate}
    \item a planet's orbital evolution due to physical processes the user wants to mimic in the simulation is well-described by time-dependant functions for the orbital parameters
    \item a parameter space exploration of chaotic evolution that ends at a desired outcome is infeasible in full n-body simulations due to computational expense, unpredictability, and the size of the parameter space
    \item the final state of a well-understood evolution process must be tuned to reach the current state of an observed system for detailed comparison with observations 
\end{enumerate}}

{An example context where criteria 1-3 are met is the well-known planetary instability model for the solar system \citep[e.g.][]{Tsiganis:2005,Levison2008,Levison2011,Nesvorny:2018,deSousa:2020}, where the rearrangement and migration of the outer planets is typically ignited by Jupiter and Saturn crossing or exiting from a mutual mean motion resonance, which causes their orbital eccentricities to be excited. The resulting planetary instability excites the orbits of the other giant planets, and they undergo a chaotic era of strong planet-planet interactions for several millions to tens of millions of years. In simulations with successful outcomes that qualitatively match the real solar system, Uranus and Neptune, characterized by their eccentric and inclined orbits, are typically damped to more circular and coplanar orbits by dynamical friction (i.e. exchange of angular momentum and energy between the planets and nearby planetesimals). At this time, Jupiter and Saturn were most likely damped to some extent as well, but not as much as the outer ice giants. The details of when an instability occurred and how long or how far Neptune migrated remain uncertain. }

{This planetary instability model is characterized by a chaotic evolution where minuscule changes to the initial parameters can produce a vastly different final configuration. Such chaos hinders the exploration of a large parameter space because the final outcomes are unpredictable. For this type of simulation, hundreds of runs are necessary just to get one that qualitatively matches the current planet configuration, making this a computationally expensive problem \citep[e.g.][]{Tsiganis:2005,Levison2008,Clement2018,deSousa:2020}. Individual control of the planets' orbital elements is useful here in allowing exploration of a larger parameter space \textit{after} the chaotic phase. For example, instability models typically include some amount of planetesimal-driven migration and dynamical friction after the instability phase to push planets toward a configuration that is more similar to the current configuration of the solar system (see Section \ref{sec:application}). Both of these physical processes are due to interactions between planetesimals and a planet, and their effect on the semimajor axis and eccentricity {of that planet} can be emulated by known functions in the simulation. 
{Because the planets are no longer having direct, scattering interactions,} we {can exert} more control to {fictitiously} evolve planets' orbits and can explore parameter space.
While it is unphysical for the orbital elements to evolve independently (eccentricity and semimajor axis are coupled, for example), fictitiously evolving planets' orbits is appropriate when the orbits' time evolution due to a physical process is known through analytical derivations and/or existing full N-body simulations. To rigorously compare these simulations with observations, the final outcome of the planets must match their current observed orbits, where the possibility of such an evolution can be assessed by comparing the particle disk configuration to the dynamical structure of TNOs. Even in simulations where planet-planet scattering is not a major driver, planets often still need an extra push to arrive at the necessary location for detailed comparisons with observations. Lastly, individual control of planet's evolution is useful when one is interested in isolating the impact of gravitational interactions between specific bodies while avoiding other interactions that are also at play. An interesting outcome of evolving planets individually is noticing that eccentricity damping of one planet can cause another planet's eccentricity to decline due to their secular coupling, as seen in \citet{Levison2011,Greenberg2013,Hansen2015}.}

In Section \ref{sec:method}, we introduce the analytical framework for adding fictitious forces and velocities to the equations of motion and then describe modifications to the n-body codes \texttt{Mercury6.2} and \texttt{REBOUND}. In Section \ref{sec:1-2planetexamples}, we show examples with one and two planets where each planet's orbital elements are evolved with arbitrary functions. In Section \ref{sec:application} we demonstrate two appropriate applications for when this code is useful. We show how we successfully reproduce  the outward migration and eccentricity damping exhibited by Uranus and Neptune in a well-known Nice model simulation \citep{Tsiganis:2005}. We also show that it is important to consider the secular coupling between Neptune and Uranus in the late-stages of orbital damping by planetesimals. {Specifically, we demonstrate that} as Neptune's eccentricity damps, Uranus' eccentricity damps indirectly as well. In Section \ref{conclusion} we summarise our results and discuss further applications for this code.

\section{Modification to The Equations of Motion}
\label{sec:method}

We aim to adjust a simulated planet's orbit such that, if no other perturbers were present in the simulation, each orbital element would independently evolve according to a specified function. The prescribed functions encapsulate the effects of processes not included in the simulation.  Note that in simulations with multiple planets, since the planet is also accelerated by other system planets, the prescribed functions do not determine the full evolution of the orbit. 
It is not possible to fully and independently evolve a planet's orbital elements using only additional forces.  
For full control, we require an update to the position of the planet at each time step (to account for changes to the velocity) as well as an update to the velocity. 
To calculate the required updates, we take partial derivatives of the equations of motion with respect to each orbital element to then form a fictitious velocity and acceleration to evolve the object given a functional form. 
This keeps the evolution of each orbital element independent from the others, which is not the case when only a user-defined acceleration is added. 
We discuss this detail further in Appendix \ref{app:accvel}.

We follow \citet{Lee2002} and expand on the work of \citet{dawson2012} and \citet{Wolff2012} to derive modifications to the equations of motion of an orbiting body to allow any form of evolution of the body's orbital elements. Modifications to \texttt{REBOUND} and \texttt{Mercury6.2} are outlined in Appendix \ref{app:derivatrives}.

\subsection{Analytical Derivation of Forces and Velocities for Arbitrary Orbital Element Evolution} \label{sec:analytical} 

The position and velocity of an orbiting body in Cartesian coordinates, ($x,y,z,\dot{x},\dot{y},\dot{z}),$ can be fully described by the six osculating orbital elements: semimajor axis ($a$), eccentricity ($e$), inclination ($i$), argument of pericenter ($\omega$), longitude of ascending node ($\Omega$), and true anomaly ($f$). In other words, $x = x(a,e,i,\omega,\Omega,f)$ and so on for the other five Cartesian coordinates.  
The total time derivative of the $x$-component of the body's velocity and acceleration can thus be written as

\begin{subequations}
    \label{eq:velacc}
    \begin{align}
\large \frac{dx}{dt} = \frac{\partial x}{\partial a}\dot{a} + \frac{\partial x}{\partial e}\dot{e} + \frac{\partial x}{\partial i}\dot{I} + \frac{\partial x}{\partial \omega}\dot{\omega} + \frac{\partial x}{\partial \Omega}\dot{\Omega} + \frac{\partial x}{\partial f}\dot{f}\\
\large \frac{d\dot{x}}{dt} = \frac{\partial \dot{x}}{\partial a}\dot{a} + \frac{\partial \dot{x}}{\partial e}\dot{e} + \frac{\partial \dot{x}}{\partial i}\dot{I} + \frac{\partial \dot{x}}{\partial \omega}\dot{\omega} + \frac{\partial \dot{x}}{\partial \Omega}\dot{\Omega} + \frac{\partial \dot{x}}{\partial f}\dot{f}
    \end{align}
\end{subequations}

\noindent and similarly for $y$, $\dot y$, $z$ and $\dot{z}$. Note that we have used $\dot{I}$ to indicate $di/dt$ for notational clarity.

In the two body problem, $a$, $e$, $i$, $\omega$, and $\Omega$ remain constant in time, and $\dot{f}$ provides the only non-zero orbital element time derivative in each of the expressions in Equation (\ref{eq:velacc}).  In other words, the central body in a two-body simulation generates evolution of the form $dx/dt = (\partial x/\partial f)\dot{f}$ and so on for the other Cartesian elements. This evolution occurs, by construction, on the timescale of an orbital period. The term $\dot{f}$ is related to the angular momentum by the equation $r^2 \dot{f} = h = na^2\sqrt{1-e^2}$, so that 

\begin{equation}
    r\dot{f} = na \frac{1+e\cos{f}}{\sqrt{1-e^2}}
    \label{rfdot}
\end{equation}

\noindent where $r = \frac{a(1-e^2)}{1+e\cos{f}}$ is the radial distance from the central body, $n = \sqrt{\frac{GM}{a^3}}$ is the mean motion, $G$ is the gravitational constant, and $M$ is the total mass of the system.

We are interested in inducing orbital element evolution on timescales much longer than the orbital period, so we separate Equations (\ref{eq:velacc}) into the first five terms, which we will evolve by hand using arbitrary functions for $\dot{a}$, $\dot{e}$, {$\dot{I}$}, $\dot{\omega}$, and $\dot{\Omega}$, and, separately, the $\dot{f}$ term. For N-body problems, the force of the central body changes $f$ by the form of Equation (\ref{rfdot}), so we do not need to prescribe an artificial evolution for this element.

Recall that the osculating orbital elements $a$, $e$, $i$, $\omega$, $\Omega$, and $f$ by definition are related to the Cartesian positions $x$, $y$, $z$ by

\begin{subequations}
    \label{eq:position}
    \begin{align}
    x &=  r\cos{\Omega}\cos{(\omega + f)} - r\cos{i}\sin{\Omega}\sin{(\omega+f)}\\
    y &= r\sin{\Omega}\cos{(\omega+f)} + r\cos{i}\cos{\Omega}\sin{(\omega+f)}\\
    z &=  r\sin{i}\sin{(\omega+f)}
    \end{align}
\end{subequations}

\noindent where $r$ is not an orbital element, but a function of three orbital elements, $r=r(a,e,f)$.
The Cartesian velocity components of the orbiting body are then the time derivative of Equation (\ref{eq:position}), allowing all elements to vary with time:

\begin{subequations}
    \label{eq:velocities}
    \begin{align} 
    \dot{x} &= \dot{r}\cos{\Omega}\cos{(\omega + f)} - r\dot{\Omega}\sin{\Omega}\cos{(\omega+f)}\\ 
    &- (r\dot{\omega} + r\dot{f})\cos{\Omega}\sin{(\omega+f)} - \dot{r}\cos{i}\sin{\Omega}\sin{(\omega+f)} \nonumber\\
    &+ r\dot{I}\sin{i}\sin{\Omega}\sin{(\omega+f)} - r\dot{\Omega}\cos{i}\cos{\Omega}\sin{(\omega + f)} \nonumber\\
    &- (r\dot{\omega}+r\dot{f})\cos{i}\sin{\Omega}\cos{(\omega+f)}\nonumber\\
    \nonumber\\
    \dot{y} &= \dot{r}\sin{\Omega}\cos{(\omega + f)} + r\dot{\Omega}\cos{\Omega}\cos{(\omega+f)}\\ 
    &- (r\dot{\omega} + r\dot{f})\sin{\Omega}\sin{(\omega+f)} + \dot{r}\cos{i}\cos{\Omega}\sin{(\omega+f)} \nonumber\\
    &- r\dot{I}\sin{i}\cos{\Omega}\sin{(\omega+f)} - r\dot{\Omega}\cos{i}\sin{\Omega}\sin{(\omega + f)} \nonumber\\
    &+ (r\dot{\omega}+r\dot{f})\cos{i}\cos{\Omega}\cos{(\omega+f)}\nonumber\\
    \nonumber\\
    \dot{z} &= \dot{r}\sin{i}\sin{(\omega + f)} + r\dot{I}\cos{i}\sin{(\omega+f)}\\ 
    &+ (r\dot{\omega} + r\dot{f})\sin{i}\cos{(\omega+f)}\nonumber
    \end{align}
\end{subequations}

\noindent where we define $\dot{r} = \frac{\partial r}{\partial a}\dot{a} + \frac{\partial r}{\partial e}\dot{e} + \frac{\partial r}{\partial f}\dot{f}$.

Recall that in the two body problem, $\dot{r}$ is derived assuming $a$ and $e$ are constant in time, producing $\dot{r}= \frac{\partial r}{\partial f}\dot{f} = \frac{na e \sin{f}}{\sqrt{1-e^2}}$ (see \citealt{SSDBook}). 

To independently control the evolution of a particle's orbital elements, the particle's position and velocity must be updated every timestep with a small additional perturbation to the momentum and force, or in other words, velocity and acceleration. 
Similar to Equation (\ref{eq:velacc}), we define a \textit{user-defined velocity} and \textit{user-defined acceleration}, which are given by

\begin{subequations}
    \label{eq:vauser}
    \begin{align}
\large \dot{x}_{\rm user} = \frac{dx}{dt} = \frac{\partial x}{\partial a}\dot{a} + \frac{\partial x}{\partial e}\dot{e} + \frac{\partial x}{\partial i}\dot{I} + \frac{\partial x}{\partial \omega}\dot{\omega} + \frac{\partial x}{\partial \Omega}\dot{\Omega}\\
\large \ddot{x}_{\rm user} = \frac{d\dot{x}}{dt} = \frac{\partial \dot{x}}{\partial a}\dot{a} + \frac{\partial \dot{x}}{\partial e}\dot{e} + \frac{\partial \dot{x}}{\partial i}\dot{I} + \frac{\partial \dot{x}}{\partial \omega}\dot{\omega} + \frac{\partial \dot{x}}{\partial \Omega}\dot{\Omega}
    \end{align}
\end{subequations}

\noindent and similarly for $\dot{y}_{\rm user}$, $\ddot{y}_{\rm user}$, $\dot{z}_{\rm user}$, and $\ddot{z}_{\rm user}$ where $\dot{a}$, $\dot{e}$, $\dot{I}$, $\dot{\omega}$, and $\dot{\Omega}$ are the arbitrary functions we define (see Section \ref{sec:1-2planetexamples}). The user-defined velocity and user-defined acceleration are added to the existing motion of the particles.
For each time step, $\Delta t$, the $x$-position and $x$-velocity of the object are updated by a term $\dot{x}_{\rm user}\Delta t$ and $\ddot{x}_{\rm user}\Delta t$, respectively and similarly for $y$ and $z$. Further discussion of our implementation in \texttt{REBOUND} and \texttt{Mercury6.2} can be found in Appendix \ref{app:mods}, including all of the partial derivatives used in Equation (\ref{eq:vauser}) (\ref{app:derivatrives}).


\section{One and two planet examples}

\label{sec:1-2planetexamples}

We demonstrate the potential of our modifications to \texttt{REBOUND} and \texttt{Mercury6.2} with example simulations consisting of one or two planets orbiting the Sun, using the \texttt{MERCURIUS} and \texttt{HYBRID} integrators for \texttt{REBOUND} and \texttt{Mercury6.2}, respectively. To compare short term and long term planetary behavior between the two N-body codes, we ran simulations with the same initial conditions, type of integrator, and output cadence. The five orbital elements, $a$, $e$, $i$, $\omega$, $\Omega$ are evolved using user-defined velocities and accelerations (Equation \ref{eq:vauser}) by defining a functional form for $\dot{a}$, $\dot{e}$, {$\dot{I}$}, $\dot{\omega}$, and $\dot{\Omega}$. For the examples in this section, we choose a combination of arbitrary functional forms for the orbital elements shown in Equation (\ref{dotaeifuncs}), where $g$ is a stand-in variable for any of the evolved orbital elements:

\begin{subequations}
\label{dotaeifuncs}
\begin{align}
    &\dot{g}(t) = \frac{\Delta g}{(t +\tau_g)}\\
    &\dot{g}(t) = \frac{2\pi\Delta g}{\tau_g}\cos{\Big(2\pi \frac{t}{\tau_g}\Big)}\\
    &\dot{g}(t) = \frac{\Delta g}{\tau_g}\exp{\Big(-\frac{t}{\tau_g}\Big)}\\
    &\dot{g}(t) = \frac{\Delta g}{\tau_g}
\end{align}
\end{subequations}

\noindent where $\Delta g$ is the scale of the change, $\tau_g$ is the timescale of the change, and $t$ is the time variable. Equation (\ref{dotaeifuncs}) corresponds to a time derivative of

\begin{subequations}
\label{aeifuncs}
\begin{align}
    &g(t) = g_0 + \Delta g\log{\Big(\frac{t}{\tau_g}+1\Big)}\\
    &g(t) = g_0 - \Delta g\sin{\Big(2\pi \frac{t}{\tau_g}\Big)}\\
    &g(t) = g_0 + \Delta g - \Delta g\exp{\Big(-\frac{t}{\tau_g}\Big)}\\
    &g(t) = g_0 + \Delta g \frac{t}{\tau_g}
\end{align}
\end{subequations}

\noindent where $g_0$ is the initial value of the parameter. In future references to Equations (\ref{dotaeifuncs}) and (\ref{aeifuncs}), the variable $g$ is replaced by the orbital element we evolve.

In the code, a user has the freedom to write any desired function to evolve an orbital parameter, which can include a nonphysical time-evolution. We expect the user to define functions that are physically motivated by full N-body simulations or analytical models. We discuss this further in Section \ref{sec:application}. The user-defined accelerations and velocities must remain small compared to the accelerations and velocities due to the other bodies in the simulation
for the integrators to produce reasonable levels of accuracy. This is a constraint due to the symplectic Wisdom and Holman integrator mappings, while implementing a non-symplectic analytical equation of motion \citep[][]{Wisdom1991,Tamayo2020}{}{}. If the additional force becomes comparable to the central gravitational force, we find that the one-planet simulations encounter large errors and eventually return \texttt{nan} values. We use the \texttt{MERCURIUS} and \texttt{HYBRID} integrators for \texttt{REBOUND} and \texttt{Mercury6.2} respectively, because they use the Wisdom-Holman scheme and switch to an adaptive, high-accuracy integrator when particles undergo close encounters. During a close encounter between two bodies, the gravitational force increases significantly and changes the orbit quickly, therefore small time-steps are essential to accurately resolve the physics.  The switch to an adaptive integrator is only triggered for close encounters however, so an additional large force and velocity, as we discuss in this paper, will not trigger the switch to an adaptive time-step, and will instead produce an inaccurate evolution.

For the single-planet example, we integrate the orbit of a Jupiter-mass planet around a solar-mass star. 
To confirm that each element is evolved independently from the others, we ran several initial tests (not shown) evolving one element at a time as well as combinations of elements.
We find that the signature of one orbital element's evolution is apparent on other elements to a negligible degree. For example, when exponentially or sinusoidally evolving only the eccentricity ($e$) of the planet,  
the semi-major axis ($a$) evolution contains a residual signature of the same functional form, though it is only noticeable at a precision of one part in $10^{7}$ for \texttt{REBOUND} and one part in $10^{4}$ for \texttt{Mercury6.2}. 
The perturbation on the semi-major axis is more pronounced when $\dot{e}$ in Equation (\ref{eq:vauser}) is large {enough for} the additional force and velocity due to the eccentricity term are no longer small compared to the velocity and acceleration of the Keplerian motion.
This confirms that the user-defined velocities and accelerations must remain small compared to the central gravitational acceleration.

\begin{figure}
    \centering
    \includegraphics[width=3.3in]{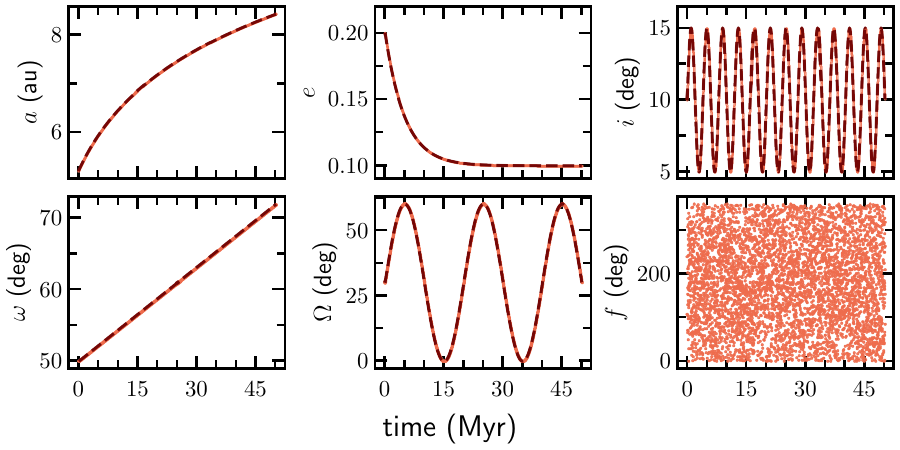}
    \caption{
    Fifty million year evolution of the semimajor axis ($a$ (au); top left), eccentricity ($e$; top middle), inclination ($i$ (deg); top right), argument of pericenter ($\omega$ (deg); bottom left), longitude of ascending node ($\Omega$ (deg); bottom middle), and true anomaly ($f$ (deg); bottom right) of a Jupiter-mass planet orbiting a solar-mass star. The planet is initialized with the orbital elements shown in Table \ref{all-sim-params}.
    The orbital elements $a$, $e$, $i$, $\omega$, and $\Omega$ are artificially time evolved by the arbitrary functions (dashed line) in Equation (\ref{aeifuncs}), denoted by the ``Eq" column in Table \ref{all-sim-params}. Each orbital element in fact evolves independently from each other and follows the functional form prescribed in the code.
    }
    \label{fig:1planetex}
\end{figure}

Figure \ref{fig:1planetex} shows an example where we artificially evolve all orbital elements with arbitrary functions for $\dot{a}$, $\dot{e}$, {$\dot{I}$}, $\dot{\omega}$, and $\dot{\Omega}$. We simulate the Jupiter-mass planet's motion for fifty million years. It is initialized {with parameters and imposed evolution shown in Table \ref{all-sim-params}.}
Through the modifications described in Section \ref{sec:method} and Appendix \ref{app:mods}, we were able to successfully evolve all elements arbitrarily and independently from each other. 

\begin{figure}
    \centering
    \includegraphics[width=4.3in]{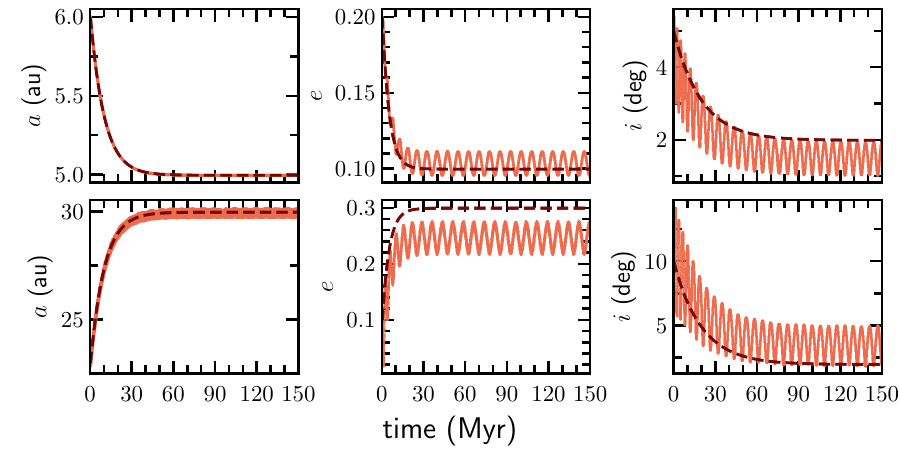}
    \caption{
    {One hundred} fifty million year evolution of the semimajor axis ($a$), eccentricity ($e$), inclination ($i$) (orange) of a Jupiter-mass (top row) and Neptune-mass (bottom row) planet orbiting a solar-mass star. The orbital elements are initialized and artificially evolved by an exponential (dark maroon dashed line) with parameters described in Table \ref{all-sim-params}. Similar to the one planet example in Figure \ref{fig:1planetex}, the additional user velocity and acceleration prescription evolves the orbital elements analogous to a desired function. However, the evolution of the eccentricity and inclination of each planet have a short timescale oscillation on top due to the secular effects between the planets. In contrast to the one planet example, where we can evolve the element to a final value, in this case, {the end result is near, but not exactly the final value. This is due to the two planets causing additional oscillations of each other's orbital elements.}
    }
    \label{fig:2planetex}
\end{figure}

\begin{figure}
    \centering
    \includegraphics[width=3.3in]{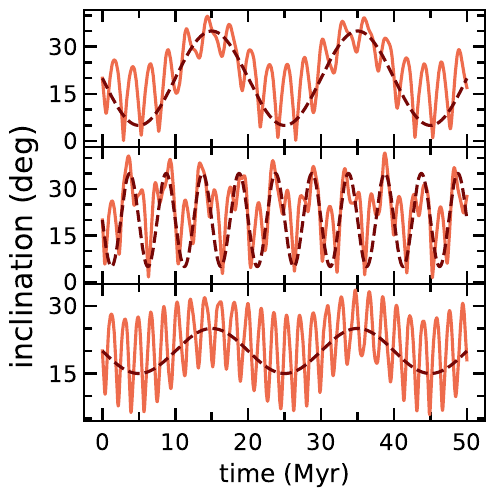}
    \caption{Fifty million year evolution of Neptune's inclination for three different simulations (solid lines) initialized as described by in Table \ref{all-sim-params}. We artificially evolve Neptune's inclination (dashed lines) as a sinusoid, varying $\Delta i$ and $\tau_i$ relative to the secular amplitude, $\mathcal{A}_{\rm sec}$ and oscillation timescale, $\tau_{\rm sec}$ with three examples: (1) $\Delta i > \mathcal{A}_{\rm sec}$ and $\tau_i > \tau_{\rm sec}$ (top panel), (2) $\Delta i > \mathcal{A}_{\rm sec}$  and $\tau_i < \tau_{\rm sec}$ (middle panel), and (3) $\Delta i < \mathcal{A}_{\rm sec}$ and $\tau_i > \tau_{\rm sec}$ (bottom panel).
    }
    \label{fig:2planet-incinations}
\end{figure}

\begin{figure}
    \centering
    \includegraphics[width=3.3in]{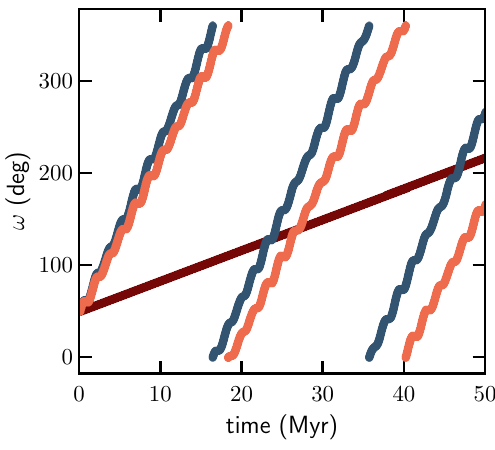}
    \caption{Two, fifty million year evolutions of Jupiter's $\omega$ under the gravitational influence of a solar-mass star and Neptune-mass planet. Both simulations are initialized with the parameters described in Table \ref{all-sim-params}. The first simulation does not include any additional velocities or accelerations (orange). In the second simulation, we add a linear perturbation (maroon) over time to Jupiter's $\omega$, making Jupiter precess faster (blue).
    }
    \label{fig:twoplanets-wOm-linear}
\end{figure}

For the two-planet example shown in Figure \ref{fig:2planetex}, we simulate a Jupiter-mass planet and Neptune-mass planet for {one hundred and} fifty million years. The planets are initialized with {values and imposed evolution shown in Table \ref{all-sim-params}. Only the semimajor axis, eccentricity, and inclination are artificially evolved.} 
{As illustrated by the oscillation in the eccentricity and inclination}, including a second planet in the system introduces gravitational forces between the planets which were not present in the one-planet simulation shown in Figure \ref{fig:1planetex}. In this three-body system, the planets secularly perturb each other, causing oscillations in the eccentricity and inclination with periods shorter than the damping timescale we prescribe. {In the one-planet example, the orbital elements follow the prescribed evolution perfectly, but that's not the case here.} Rather than having full control of the final value of a given element, the scale of change, $\Delta g$ (i.e. final value minus initial value), in Equation (\ref{aeifuncs}c) sets an \textit{approximate} final outcome in simulations with more than one planet. Jupiter's secular effect on Neptune is greater than Neptune's effect such that Neptune often over- or under-shoots the prescribed final eccentricity by an amount comparable to the amplitude of the secular oscillation. 
{We find that when prescribing an exponential growth or decay for eccentricity or inclination, the simulated average $e$ or $i$ may asymptote toward a value that is greater or smaller than the specified final value, as in Figure \ref{fig:2planetex}. In analyzing hundreds of simulations, we did not find an obvious trend for how simulation conditions affect the final simulation $e$ or $i$ value relative to the prescribed final value.} 

While the average simulation $e$ or $i$ value may asymptote below or above the value we prescribe, {this is not a major issue} because the simulation is still evolving accurately under the conditions provided. This could, however, be an issue if the user's primary goal is to reach a specific average value. Damping to an exact value will not happen each time in the 2+ planet simulations because the gravitational perturbations between the planets may dominate {under certain conditions}. This interdependence is important because we are interested in modeling how planets behave under physical (absent from the simulation) processes in addition to planet interactions. 
We note that because we are using additional accelerations and velocities to model the behavior of an absent physical process, we are not conserving the angular momentum or energy of the planet. The energy and angular momentum gained or lost by the planet would be lost or gained by the population of planetesimals (or other bodies) that the fictitious forces and velocities are mimicking.

The simulation setup in Figure \ref{fig:2planetex} is such that Neptune's secular inclination amplitude is large since Jupiter started at a high inclination. The secular eccentricity perturbations for both Jupiter and Neptune are small compared to the eccentricity user-defined perturbations (i.e. the total user input change is several times larger than the secular amplitude), whereas, the exponential change for the inclination is on the same order as the amplitude of the secular oscillations.  
Understanding how user-defined perturbations compare to the gravitational perturbations between planets is essential for running effective models and knowing the limit of control one has on the evolution of planets. 

We highlight {the limitation for controlling planets' evolution} further in Figure \ref{fig:2planet-incinations} with three {different simulations. Each simulation includes the star and two planets with parameters defined in Table \ref{all-sim-params}. We apply a sinusoidal evolution to both planets' inclination to illustrate how the evolution differs for} different values of $\Delta i$ and $\tau_i$ relative to the secular amplitude, $\mathcal{A}_{\rm sec}$ and secular oscillation timescale (1/frequency), $\tau_{\rm sec}$. {Figure \ref{fig:2planet-incinations} shows the Neptune-mass planet's inclination evolution over fifty million years.}
When $\tau_i > \tau_{\rm sec}$, the long-term evolution is dominated by the user-defined forces. When $\Delta i > \mathcal{A}_{\rm sec}$, the prescribed sinusoidal evolution is noticeable, {as} demonstrated in the top panel of Figure \ref{fig:2planet-incinations}. The next two simulations show the case where $\tau_i < \tau_{\rm sec}$ and $\Delta i < \mathcal{A}_{\rm sec}$, respectively. 
When $\Delta i < \mathcal{A}_{\rm sec}$, the secular oscillation frequency is equivalent to the case when no perturbations are present.
Note, when the amplitude of the induced inclination oscillation is higher (top panel; Figure \ref{fig:2planet-incinations}), the secular oscillation frequency is lower compared to the case with a small amplitude imposed oscillation (bottom panel). The secular frequency is altered because the imposed evolution is sinusoidal; we are effectively adding a frequency that interferes with the natural secular oscillation frequency. If we instead impose an exponential function, which is more typical for damping and exciting eccentricities and inclinations, the secular frequency does not substantially change (not shown).

The artificial evolution of $\omega$ and $\Omega$ in a multi-planet case requires a more careful consideration than that of the other elements. One cannot choose completely arbitrary functions for their evolution because the interactions between the planets dominate many attempts to control them. If planets are not strongly coupled, then $\omega$ and $\Omega$ will be easier to force, but this is typically not the case in a multi-planet system. For example, for several test simulations with Jupiter and Neptune, we added and subtracted a small amount of precession to Jupiter by imposing a linear evolution on top of the {$\omega_J$}. This is helpful if one desires to slow down or speed up the precession of a planet. In these simulations, Neptune's {$\omega_N$ remained the same, and $\varpi_J - \varpi_N$ remained constant}, {where $\varpi = \Omega + \omega$}. In another test, we increased {$\omega_J$ and decreased $\omega_N$ so that $\varpi_J - \varpi_N$ changed over time}. For this test, the amplitude and frequency of the eccentricity and inclination of each planet varied over time, thus changing the secular structure. One example of increasing Jupiter's $\omega_J$ is shown in Figure \ref{fig:twoplanets-wOm-linear}.
Fictitious evolution of $\omega$ and $\Omega$ is worth considering for scenarios where the user is interested in altering the interactions between the planets, for example in order to track how varying precession affects thousands of particles in the disk. We discuss these implications further in Section \ref{conclusion}.

\section{Applications for Artificial Planetary Evolution}
\label{sec:application}

In this section we show two example applications of artificial planet evolution: (1) reproducing phases of migration and eccentricity damping {for Uranus and Neptune} \textit{after} a planetary instability (Section \ref{sec:tsiganis}) and (2) investigating the secular coupling between Uranus and Neptune by artificially damping Neptune's eccentricity and seeing the effect on Uranus' eccentricity (Section \ref{sec:neptune_damp_uranus}). For both applications, our initial conditions are set by the planet parameters after a chaotic evolution and before a {period of smoother} migration from the simulation described in Figure 1 of \citet{Tsiganis:2005}. 

\subsection{Reproducing Epochs of Migration and Eccentricity Damping}
\label{sec:tsiganis}

Reproducing epochs of migration and eccentricity damping after an instability phase is important for studying how the small bodies in the outer solar system—TNOs in mean motion resonance, in particular—are affected by various types of planet evolutions post-instability. In this paper we focus on reproducing the planet evolution and apply this to TNO populations in future work. 
We aim to reproduce two smooth migration and eccentricity damping phases of Uranus and Neptune in the well-known planet instability model from \citet{Tsiganis:2005} to show the usage of this code. We note that it is now well understood that the simulation in \citet{Tsiganis:2005} was not able to reproduce all aspects of the planets' orbits. We use it here as a test-case for validation of our code because it has been well studied. 

{First, we describe the two phases of semi-major axis and eccentricity evolution from \citet{Tsiganis:2005} when fit by an exponential function (Section \ref{sec:bestfit_eccentricity}). Next, we calculate the eccentricity damping timescale due to analytical dynamical friction theory and compare it with the simulation, best-fit timescale to confirm that dynamical friction is the dominant force at play (Section \ref{sec:dynamical_friction}). Lastly, we present two simulations where we artificially evolve the planets' semimajor axis and eccentricity and show that we reproduce migration and eccentricity damping phases from \citet{Tsiganis:2005} (Section \ref{sec:tsiganis_simulation}).} 
With this example, we will demonstrate how one can artificially evolve planets according to a physically motivated process. 

\begin{figure}
    \includegraphics[width=3.3in]{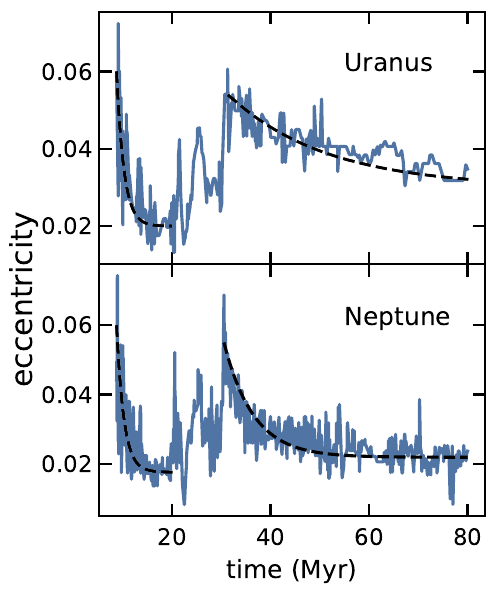}
    \caption{Eccentricity for Uranus (top panel) and Neptune (bottom panel) calculated by using the semimajor axis, pericenter, and apocenter curves from Figure 1 of \citet{Tsiganis:2005}. The best-fit exponential damping timescale at the early stage (before 20 Myr) is 1.5 Myr for both Uranus and Neptune. At late times (after 30 Myr), the best-fit exponential timescales are 20 Myr and 6 Myr for Uranus and Neptune, respectively {(see Section \ref{sec:bestfit_eccentricity} for the best fit final and initial eccentricities)}.  
    }
    \label{fig:extrapolate}
\end{figure}

\subsubsection{Finding the Best-fit Exponential Migration and Eccentricity Damping Timescales}
\label{sec:bestfit_eccentricity}

We extrapolate the {semi-major axis, apocenter, and pericenter} evolution of the giant planets from Figure 1 in \citet{Tsiganis:2005} {for the time range between 8.6 Myr and 80 Myr}.\footnote{We use the online tool \texttt{Webplot Digitizer} to extract the data from the figure (\hyperlink{https://apps.automeris.io/wpd4/}{https://apps.automeris.io/wpd4/}).} We calculate ``early" (8.6-20 Myr) and ``late" (30-80 Myr) eccentricity time-evolution profiles for Uranus and Neptune using the values in the three curves since $a_{\rm peri} = a(1-e)$ and $a_{\rm apo}=a(1+e)$ (see Figure \ref{fig:extrapolate}). 
We use {\texttt{scipy.optimize.curve\_fit} to} fit an exponential function {(Equation \ref{aeifuncs}c)} to the data to find best-fit {parameters for semi-major axis and eccentricity.} 
{Because the fit is done with data points that were extracted by hand, which are not evenly spaced in time, this fit is {just meant to be an estimate of the overall motion}. For Uranus' early and late phases, we find ($e_0 = 0.06$, $e_f = 0.018$, $\tau_e = 1.5$ Myr; $a_0 = 18.6$ au, $a_f = 19.3$ au, $\tau_a = 7.2$ Myr) and ($e_0 = 0.054$, $e_f = 0.03$, $\tau_e = 20$ Myr; $a_0 = 19.3$, $a_f = 19.7$, $\tau_a = 17$ Myr), respectively. For Neptune's early and late damping phases, we find ($e_0 = 0.06$, $e_f = 0.02$, $\tau_e = 1.5$ Myr; $a_0 = 23.7$ au, $a_f = 25.4$ au, $\tau_a = 11$ Myr) and ($e_0 = 0.055$, $e_f = 0.022$, $\tau_e = 6$ Myr; $a_0 = 25.5$ au, $a_f = 26.43$ au, $\tau_a = 9.6$ Myr), respectively.}

{In this simulation from \citet{Tsiganis:2005}, massive planetesimals are present, and have a large role in damping the planets' eccentricities through dynamical friction. The damping timescale and amount of damping due to dynamical friction depend on the mass density of the planetesimal disk (i.e. a larger mass in planetesimals per volume produces more momentum exchange between a planet and its neighboring planetesimals, and the planet will damp faster and further). At early times, the mass density across the planetesimal disk is higher than late times and likely comparable at the locations of Uranus and Neptune, so it is expected that they have similar, short damping timescales in this phase (see below). After continuous strong gravitational scattering events, the mass density decreases; whether this change is uniform across the disk in the simulation from \citet{Tsiganis:2005} is not {explicitly stated}. It is possible that the local mass density near Neptune is larger than that around Uranus and as a result, Neptune's eccentricity is damped further and faster than Uranus'. We investigate this point by running a few dynamical instability simulations and track how the number density across the disk evolves over time. We found that for the first $\sim$10 Myr after a dynamical instability, the number density remains approximately uniform across the locations of Uranus and Neptune, consistent with the similar damping times evident at early times in Figure \ref{fig:extrapolate}. 
At later times, the region near Uranus eventually becomes more cleared, but we cannot verify the timescale of that clearing in the \citet{Tsiganis:2005} simulation. 
We note that it is possible that the secular coupling between Uranus and Neptune has a role in this late damping phase. We discuss this point further in Section \ref{sec:neptune_damp_uranus}.}

\subsubsection{Calculating Eccentricity Damping Timescale due to Dynamical Friction}
\label{sec:dynamical_friction}

To calculate an eccentricity damping timescale due to dynamical friction we use the parameters of the disk stated in \citet{Tsiganis:2005} and in the caption of their Figure 1. The information that is not explicitly stated is inferred to the best of our ability and does not impact the timescale substantially. The initial orbits of Jupiter, Saturn, Uranus, and Neptune are nearly circular and coplanar ($e\sim i\sim 0.001$) with initial semimajor axes 5.45, 8.65, $\approx$17, and $\approx$12, respectively. In that simulation, Uranus and Neptune switch orbital positions, a phenomenon that occurred $\sim 50$\% of the time. Their simulated planetesimal disk is 35 $M_{\rm E}$, composed of 3500 equal-mass particles up to 30 au, and has a surface density that falls linearly with heliocentric distance. We assume an inner edge of 17 au, {since the outer planet is initially located at this distance}. All particles are initialized with ($e \approx \sin i \approx 0.05$). These values are essential in calculating the timescale at which Neptune would damp due to dynamical friction, since the amount of momentum exchange between the planets and the planetesimals in the disk depends on the mass density of the medium.

{As a planet travels within a sea of planetesimals, the random velocities of the planet, $v_P$, and the planetesimals, $v_l$, (i.e. the velocity deviating from the orbital velocity of a circular motion) evolve through dynamical interactions between the bodies. The subscripts $l$ and $P$ represent the planetesimal and planet, respectively. We are in the regime where the planet is much larger than the planetesimals so that $M_P v_p > m_l v_l$. As a result, the big body must collide with many small bodies before its velocity changes significantly. It is also the case that $v_{P,\rm esc} > v_l > v_{P,\rm H}$, where $v_{P, \rm esc}$ is the escape velocity from the planet's surface and $v_{P,\rm H}$ is the planet's Hill velocity.  As illustrated in \citep{Goldreich2004}, this means that the primary exchange of momentum between the planet and planetesimals is by collisionless gravitational deflections, enhanced by gravitational focusing. Over time, the exchange of momentum leads to dynamical friction, which damps the planet's eccentricity (see \citealt{Goldreich2004} for a review). The planet will stop damping once equipartition of random kinetic energies is reached (i.e. $M_P v_P^2 = m_l v_l^2$, which we don't expect to occur given that $M_p \gg m_l$) or once most of the planetesimals have been ejected. 
Encounters between the planet and planetesimals also change the relative velocity between the two bodies, $v_{\rm rel} \sim \sqrt{v_{l}^2 + v_{P}^2}$, which can be dominated by the planet's or planetesimal's random velocity. 
If the planet's eccentricity is high enough, $v_P > v_l$, the dynamical friction formulation is slightly different than the case where planetesimals' eccentricities are more excited and $v_l > v_P$. Given that our calculations are meant to model dynamical friction in a dynamical upheaval scenario, we account for both cases.} 

The order of magnitude momentum exchange per time {due to encounters between the planet and planetesimals takes the form}  $M_P\frac{dv}{dt} = - (m_l v_{\rm rel}) n \sigma v_{\rm rel}$, where $mv$ is the momentum transfer per encounter and $n\sigma v$ is the number of encounters per time.  Here, $M_P$ is the mass of the big body, $m_l$ is the mass of a small body, $n$ is the number density (so that  $\rho = m_ln$ is the mass  density) of the {planetesimal} disk, and $\sigma$ is the cross sectional area for the encounters. 
{
Gravitational focusing is important, so the cross section, $\sigma = \pi b^2$, is determined by the impact parameter $b \approx GM_P/v_{\rm rel}^2$.}

{When $v_P > v_l$, the planet is plowing through a planetesimal disk and encounters are clustered in front of the planet so that $v_{\rm rel} \approx v_P$.   The momentum exchange per strong scattering encounter is $\sim$$m_lv_P$, and the momentum exchange formula becomes $\frac{dv_P}{dt} = -\frac{2\pi \rho G^2M_P}{v_{P}^2}\ln(1+\Lambda^2)$, which matches that in \citet{Ford2007}\footnote{The calculation in \citet{Ford2007} assumes the planet interacts with the disk twice per orbit since they assume $i_P \gg i_l$, whereas we are assuming that the planet and planetesimal disk have similar inclinations (i.e. the planet is embedded in the disk and feels the perturbation on behalf of the disk for the duration of an orbital period). Using their notation, our $\Delta t$ is the full orbital period rather than the amount of time spent crossing the disk per orbit, $\sim$$H/|v_z|$, where $v_z$ is the vertical component of the planet's velocity.} and \citet{Binney2008}.  Here, the Coulomb parameter, $\Lambda = b_{\rm max}/b_{\rm min}$, accounts for the contribution of encounters at various impact parameters.}

{When $v_P < v_l$, we need to account for momentum exchange in front and behind the planet. This results in $M_P\frac{dv_P}{dt} \approx n \sigma (v_l - v_P) m_l (v_l - v_P) - n \sigma (v_l + v_P) m_l (v_l + v_P) 
\sim -\rho \Big(\frac{GM_P}{v_{\rm rel}^2}\Big)^2 v_l v_P$. This form matches that found in \citet{Goldreich2004}.
Using  $v_{\rm rel} \approx v_l$,
the dynamical friction force in this regime is thus given by}

\begin{equation}
    \frac{dv_P}{dt} = -\frac{2\pi G^2M_P\rho}{v_{l}^3}v_P\ln(1+\Lambda^2)
    \label{eq:dvP_dt}
\end{equation}

\noindent {We approximate the maximum and minimum impact parameters as $b_{\rm max} = R_H + H$ and $b_{\rm min} = GM_P/v_l^2$, where $R_H = a_P(M_P/3M_*)^{1/3}$ is the Hill sphere radius, $H=v_l /\tilde{\Omega}$ is the scale height of the disk, and $\tilde{\Omega}$ is the orbital angular frequency \citep{Stewart2000}.}

\noindent{
In equipartition, $\boldsymbol{v}$ has two degrees of freedom parallel to the plane and one degree of freedom perpendicular to the plane, so that $e \approx \sqrt{2}\sin{i}$. We approximate $v_P = \sqrt{3/2}e_P \tilde{\Omega}_P a_P$ and $v_l = \sqrt{3/2}e_l\tilde{\Omega}_P a_P$, where we have made the simplifying assumption that the semi-major axes and hence orbital angular frequencies of the planetesimals and planet are the same. Plugging in for the planetesimal and planet random velocities,
Equation (\ref{eq:dvP_dt}) simplifies to
}

\begin{equation}
    \frac{de_{P}}{dt} = -\frac{4\pi}{3\sqrt{3}} \left(\frac{\Sigma a_P^2 M_P}{M_*^2}\right)\left(\frac{e_P}{e_{l}^4}\right)\tilde{\Omega}_P\ln(1+\Lambda^2)
    \label{eq:deP_dt}
\end{equation}

\noindent {where $\Sigma= (m_{tot}/(2\pi r_{out}(r_{out}-r_{in}))(r_{out}/a_P)$, $\rho = \Sigma/(2H)$, and $H = a_P \sin{i_l} \approx a_P e_l/\sqrt{2}$. The characteristic order-of-magnitude damping timescale due to dynamical friction is then }
\begin{equation}
    t_{\rm damp} = \frac{e_{P}}{de_{P}/dt}.
    \label{eq:tdamp}
\end{equation}

\noindent {where $de_P/dt$ is evaluated using Equation (\ref{eq:deP_dt}).}

For Neptune we plug in $M_P=M_{\rm N}$, $a_P=25.5$ au for the early ($e_P=0.06$) and late ($e_P=0.055$) damping phases. {For Uranus, we plug in $M_P=M_{\rm U}$, $a_P=19.3$ au for the early ($e_P=0.06$) and late ($e_P=0.054$) phases. For both cases, we plug in $M_*=M_\odot$, $e_l=e_P$, $m_{tot}= 35 M_{\oplus}$, $r_{out}=30$ au, and $r_{in}=18$ au. Then, we scale the mass density, $\rho = \Sigma/(2H)$, to account for the dispersal after millions of years by dividing $\rho$ by 50 and 500 for the early and late damping phases, respectively. We calculate
$\approx0.8$ Myr and $\approx 6.1$ Myr for Neptune's early and late damping phases, respectively. For Uranus, we calculate $\approx 0.7$ Myr and $\approx 5.5$ Myr.}
These values match the damping timescales from the simulation within a factor of 2 (see Figure \ref{fig:extrapolate}) {with the exception of the late damping phase of Uranus, possibly suggesting that another process is in effect here (see Section \ref{sec:neptune_damp_uranus})}.
{To estimate the normalizing factor for the mass density, which accounts for the disk dissipation over time, we tracked how the number density near Neptune evolved in an example dynamical upheaval simulation. In comparing the calculated damping timescale with the best-fit simulation timescale, we find that} the physical process of dynamical friction is responsible for damping Neptune's eccentricity in the simulation, and it turns out to be well-represented by an exponential function with an order of magnitude estimated timescale. 

{ We don't reproduce the final asymptotic eccentricities with this formulation. Without the mass density of the disk as a function of time, we are unable to integrate Equation (\ref{eq:deP_dt}). Uranus' final eccentricity is higher than Neptune's, possibly because the mass density around it dissipated faster, but the simulation data needed to test this hypothesis is unavailable. The planets' eccentricities eventually asymptote because the overall disk's mass density decreases significantly over time.}

\begin{figure}
    \centering
    \includegraphics[width=3.3in]{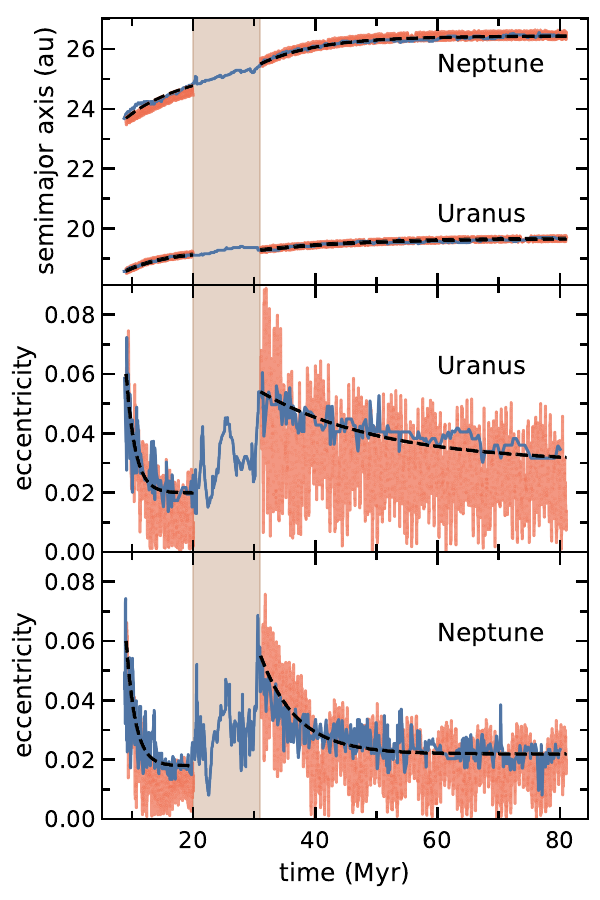}
    \caption{Two example simulations (orange) showing the evolution of the semimajor axes (top panel) and eccentricities of Uranus (middle panel) and Neptune (bottom panel) representing an early and late damping phase (separated by the shaded region). {Initial conditions and imposed evolution are found in Table \ref{all-sim-params}.} We artificially evolve the two outer planets' orbits using the best-fit parameters {discussed in Section \ref{sec:bestfit_eccentricity}}. The evolution from Figure \ref{fig:extrapolate} is shown for reference in blue.
    The dashed curves show the best fit exponential evolution for each early and late smooth migration and eccentricity damping phases. 
    }
    \label{fig:tsiganisearlymig}
\end{figure} 

\subsubsection{Simulating the Early and Late Migration and Eccentricity Damping Phases}
\label{sec:tsiganis_simulation}

{Having confirmed} that the use of an exponential decay function to represent the dynamical friction process is appropriate, we use the best-fit values {described in Section \ref{sec:bestfit_eccentricity}} to {artificially} migrate and damp {Uranus and Neptune}. In Figure \ref{fig:tsiganisearlymig} we show a simulation where we successfully reproduce early and late epochs of outward migration (top panel) and eccentricity damping for Uranus (middle panel) and Neptune (bottom panel). {The initial planet parameters and imposed evolution are shown in Table \ref{all-sim-params}.} {We note that because this is a chaotic problem, slightly different initial conditions produce vastly different results. For example, a minuscule change in argument of pericenter may enact a different gravitational interaction between the planets that causes one planet to be ejected from the system. We ran 40 instances of these simulations with the same parameters as described in Figure \ref{fig:tsiganisearlymig}, except the phase angles ($\omega$, $\Omega$, and $f$) were initialized randomly from a range of 0 to $2\pi$. We saw two different outcomes: in 20 simulations, the planets behaved similarly to those in Figure \ref{fig:tsiganisearlymig}, and in the remaining simulations, the planets underwent a chaotic evolution where Uranus and/or Neptune reached high semi-major axis/eccentricities or were ejected.}

With this exercise, we show an appropriate use for aritficially evolving planets' orbits, motivated by a physical process. While for this case, our goal was to reproduce the scenario in a specific simulation, for the case of outer solar system models, the goal is to evolve the giant planets' orbits such that they match the current observed state of the Solar System. {To do this, one would use a function and parameters that represents a realistic temporal evolution (see Equation \ref{aeifuncs}).}

\subsection{Effect of Damping Neptune on Uranus}
\label{sec:neptune_damp_uranus}
An interesting outcome of our code tests is that of isolating the impact of exclusively damping Neptune's eccentricity as if it had been damped by dynamical friction. With this method, we study how the secular coupling between Uranus and Neptune at early and late stages affects the evolution of Uranus, {when only artificially evolving Neptune}. This effect is similar to the redistribution of tidal damping by secular coupling between planets \citep{Greenberg2013}. By modeling Neptune's evolution artificially and without a massive planetesimal disk, we can study the interactions between Uranus and Neptune in isolation. 

We find that as Neptune's eccentricity damps, Neptune exerts a force on Uranus, such that Uranus' eccentricity damps as well. The magnitude of the effect depends on a few parameters. If Neptune's damping timescale is short ($\lesssim$ 1 Myr) or the change in eccentricity is small ($\lesssim$ 0.01), then the effect on Uranus is negligible. {We run simulations with a variety of initial eccentricities for Uranus and Neptune and damping timescales to understand this effect. All simulations contain the 4 giant planets with initial parameters and imposed evolution described in Table \ref{all-sim-params}. We show three examples in the following text with parameters described in Table \ref{all-sim-params}. Similar to the simulations described in Figure \ref{fig:tsiganisearlymig}, we ran 40 simulation instances of each example, where we randomly chose the phase angles ($\omega$, $\Omega$, $f$) between $0$ and $2\pi$. } 

\begin{figure}
    \centering
    \includegraphics[width=3.3in]{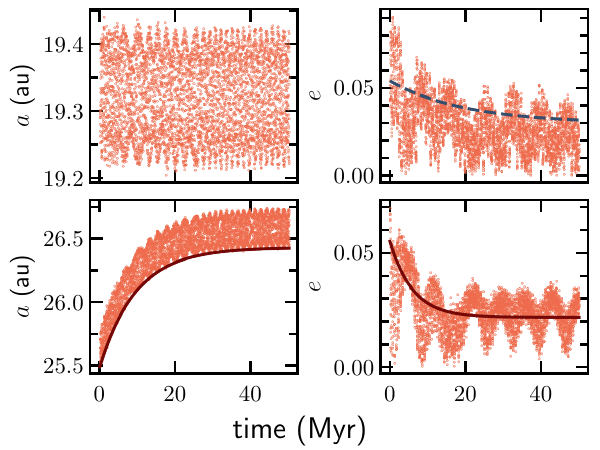}
    \caption{Fifty million year simulation {(orange dots)} initialized with the same parameters as those in the late stage evolution from Figure \ref{fig:tsiganisearlymig} {(also, see Table \ref{all-sim-params})} and only showing Uranus (top row) and Neptune (bottom row). Neptune migrates and its eccentricity is damped with the functional form shown in maroon (bottom panels). Uranus' is not artificially damped in this simulation, rather, it is indirectly damped (top {right} panel). The dashed line shows the same damping timescale, 20 Myr, for Uranus as in Figure \ref{fig:tsiganisearlymig} to show that the indirect damping of Uranus is comparable to that from \citet{Tsiganis:2005}.}
    \label{fig:uranusdampmore}
\end{figure}

Figure \ref{fig:uranusdampmore} shows a simulation that is initialized in the same way as {the late-phase simulation from} Figure \ref{fig:tsiganisearlymig}, but only Neptune's eccentricity and semimajor axis are {artificially} evolved (bottom panels; maroon lines). {We keep Uranus' semi-major axis fixed since it only migrated a fraction of an au, and its migration does not affect the eccentricity evolution here. Neptune is expected to migrate a larger distance, and so we represent that in these simulations. Through experiments, we found that this coupling behavior is independent of the migration of both planets. We find that} Uranus' eccentricity is indirectly damped in this simulation through interactions with Neptune (top right panel). {Uranus damps to a similar extent to the evolutionary behavior seen in the simulations from \citet{Tsiganis:2005}. For reference, we plot the same exponential damping curve from Figure \ref{fig:tsiganisearlymig} on top ($\tau_e = 20 Myr$, $e_0 = 0.054$, $\Delta e = -0.024$). In the 40 simulations we ran, we found two outcomes: $12$ simulations behaved similarly to Figure \ref{fig:uranusdampmore} and the remaining simulations showed chaotic planet evolution.}

\begin{figure}
    \centering
    \includegraphics[width=3.3in]{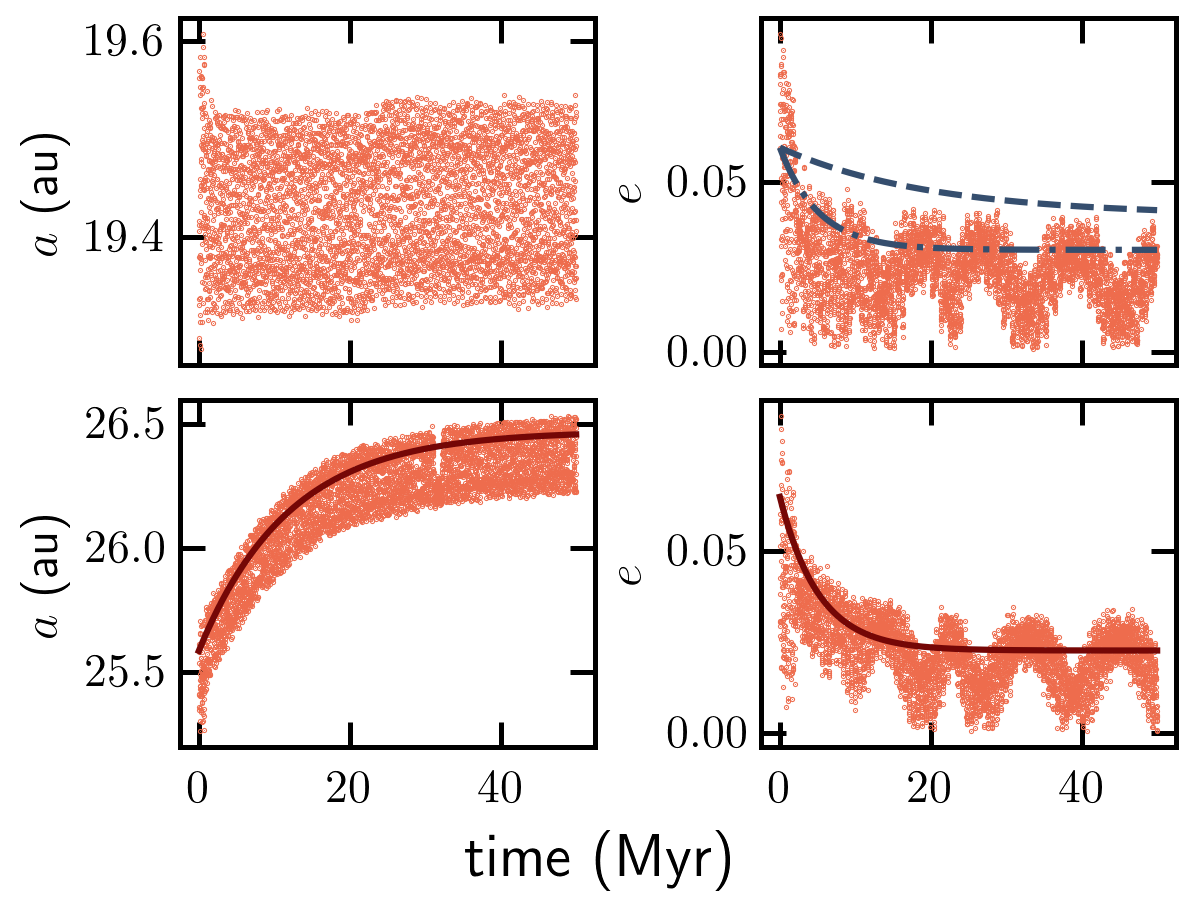}
    \caption{Fifty million year simulation similar to that in Figure \ref{fig:uranusdampmore}, except Uranus and Neptune have higher initial eccentricities of 0.06 and 0.065 and Neptune's eccentricity is damped with a timescale of 5.1 Myr. For reference two dashed lines are plotted on top of Uranus' eccentricity (top right) with $\tau_e = 20$ Myr and $\tau_e = 5.1$ Myr. 
    }
    \label{fig:highUN_edamp}
\end{figure}

We ran additional simulations similar to those shown in Figure \ref{fig:uranusdampmore}, but we vary the initial eccentricity of Neptune and Uranus to investigate how {a different} $\Delta e$ {and $\tau_e$} will affect the indirect damping on Uranus. {We show one example in Figure \ref{fig:highUN_edamp}. If the initial eccentricity of both Uranus and Neptune are larger ($0.06$ and $0.065$, respectively) and Neptune's damping timescale is shorter ($\tau_e=5.1$ Myr), Uranus' eccentricity damps a larger amount and faster than in the example from Figure \ref{fig:uranusdampmore}. We saw two outcomes in the 40 simulation instances: $12$ simulations behaved similarly to Figure \ref{fig:highUN_edamp} and the remaining simulations showed chaotic planet evolution.} This effect is interesting to follow up on when studying the late stages of planetary evolution in our solar system {where the exact details of the planets' eccentricities and planetesimal remain uncertain}. {In a simulation representing the early phase described in Figure \ref{fig:tsiganisearlymig}, we only exponentially damped Neptune's eccentricity from 0.06 to 0.02, where $\tau_e = 1.5$ Myr and saw no effect on Uranus (plot not shown). The damping timescale is too short at this early phase (i.e., it is comparable to and shorter than the secular frequency of Uranus and Neptune, respectively), so Neptune's eccentricity damping does not indirectly damp Uranus.} The secular eccentricity timescales of Uranus and Neptune are 0.5 Myr and 2 Myr, respectively \citep[see, e.g.][]{SSDBook}; therefore, the secular coupling between Uranus and Neptune, that would cause Neptune to damp Uranus' eccentricity indirectly, is stronger at late times when the damping timescale is $\ga 6$ Myr.

\begin{figure}
    \centering
    \includegraphics[width=3.3in]{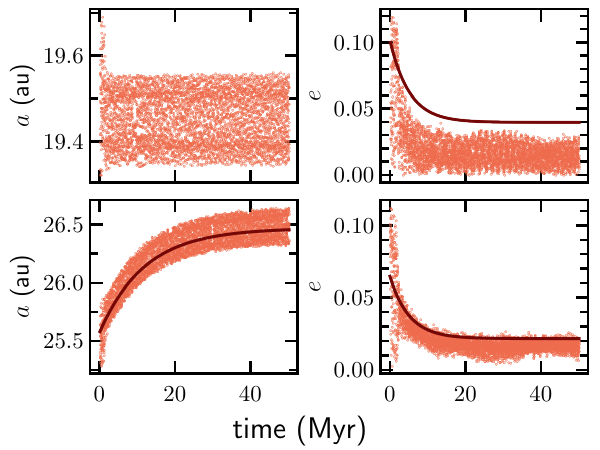}
    \caption{Fifty million year simulation (orange dots) initialized with the same parameters as those in the late stage evolution from Figure \ref{fig:tsiganisearlymig}, except Uranus (top row) and Neptune (bottom row) start at eccentricities 0.1 and 0.065, respectively. {Neptune is migrated artificially (maroon line, bottom left panel) to 26.48 au with an exponential timescale of 11.9 million years. Both planets' eccentricities are artificially damped with the same timescale of 5 Myr (solid maroon line).} Uranus' eccentricity damps further than the amount we prescribe {(top right panel)} due to the secular effects with Neptune.}
    \label{fig:uranus0.1e}
\end{figure}

While the giant planet instability model has reproduced many important features of the currently observed structure within the giant planets and TNOs, the evolution in between an instability and the current TNO configuration remains unknown. This includes the possibility that Uranus and Neptune were launched onto orbits with eccentricities as high as 0.4 and damped at various magnitudes and rates depending on the mass density of the planetesimal disk. With this in mind, we run a last suite of simulations similar to those described in Figure \ref{fig:uranusdampmore} and \ref{fig:highUN_edamp} but with Uranus' eccentricity initialized at values higher than Neptune's. One such example is shown in Figure \ref{fig:uranus0.1e}. In this simulation, Uranus' and Neptune's eccentricity are initialized at 0.1 and 0.065, respectively, representing a scenario just before a damping phase. Both planets are artificially damped with the same eccentricity-damping timescale {of 5 Myr, but Neptune is forced to damp to 0.023 and Uranus to 0.04. Prescribing the same damping timescale is motivated by a few of our tests,} instability simulations that show that the planetesimal number density near Uranus and Neptune is comparable after a few million years post-instability (not shown). In this simulation, Neptune's effect on Uranus is significant enough to damp Uranus' eccentricity {to a midpoint below 0.02, which is lower than we prescribed} (top right panel). {We saw three outcomes in the 40 simulations instances we ran: 7 simulations behaved similarly to Figure \ref{fig:uranus0.1e}, 3 simulations showed signs of planet interactions and Uranus' eccentricity ended up near the final eccentricity we prescribed, and the remaining simulations showed chaotic planet evolution. Both of the simulations shown in Figure \ref{fig:highUN_edamp} and \ref{fig:uranus0.1e} show a quick, jumping feature in the semi-major axis and eccentricity at early times, which can be attributed to Uranus and Neptune crossing a mean motion resonance.}

\section{Discusion and Summary}
\label{conclusion}

The modifications to \texttt{REBOUND} and \texttt{Mercury6.2} we present in this paper are helpful for running suites of N-body simulations that can probe a large parameter space of orbital configurations for the purpose of constraining the evolution of the outer solar system. We demonstrate how to evolve particles' orbital evolutions with arbitrary functions {by applying user-defined velocities along with the more common, user-defined accelerations}. We show that with a one-planet simulation, all orbital elements can be evolved, {as prescribed and} independently from each other. Introducing a second planet to the system, adds new gravitational interactions between the planets, which affects the amount of control one has over the artificial evolution of each orbital element. Secular perturbations between the planets produce oscillations in eccentricity and inclination, {so the final simulated average $e$ and/or $i$ may be lower or higher than the prescribed final value. Planets may also cross mutual mean motion resonances, inducing jumps which may dominate the orbital behavior over the artificial evolution.} It is {therefore, important to carefully choose initial conditions and} consider how the secular frequencies and amplitudes of the planets compare to the desired scale of change and timescale of change to successfully evolve $e$ and $i$ accordingly. {In our simulations, we do not impose the conservation of angular momentum, which is nonphysical, but our primary goal is to mimic the presence of physical processes absent from the simulation (e.g. a planetesimal disk interacting with planets). By evolving the planets' orbits according to a physically motivated process and keeping the gravitational interactions between planets, we allow ourselves to have more control in the simulation while maintaining the effect of planet interactions.}

We demonstrate appropriate uses of this code, emphasizing the need for physically motivated functional forms for a given orbital evolution. {We focus on Uranus' and Neptune's artificial orbital evolution in the context of reproducing epochs of smoother migration and eccentricity damping phases in a well-known planetary instability simulation \citep{Tsiganis:2005}.} We choose an exponential function to evolve these parameters and confirm eccentricity damping timescales with dynamical friction derivations appropriate for the planetesimal disk in their simulation. We are able to reproduce two stages of {smoother} migration and eccentricity damping. {Artificial evolution is also useful for outer solar system studies that probe how different giant planet dynamical histories affect the orbital configuration of TNOs. A variety of proposed migration histories within the outer solar system, modeled through numerical simulations, produce different planet and TNO structure \citep[for reviews see][]{Nesvorny:2018,Gladman:2021}. In fact, a similar migration history, with minuscule changes in initial conditions, could also produce vastly distinct planet trajectories as is the case for planetary instability numerical models as discussed in Section \ref{sec:application}, where chaos is the underlying driver \citep[e.g. see][for a discussion on the expense of running such simulations]{volk2019}.} {A realistic portrayal of a planetary system in an N-body simulation would require all solar system planets and billions of massive planetesimals. The chaotic nature of this problem is further accentuated by the additional gravitational interactions between each body, making exploration of a large parameter space of initial conditions computationally infeasible. The ability to artificially evolve planets' orbits is therefore beneficial for investigating how TNOs in mean motion resonance (MMR) with Neptune evolve as a result of various migration histories. }

{An interesting application of this numerical approach is the ability to} isolate the impact of one planet's eccentricity damping {on another planet,} in the absence of a planetesimal disk. We find that artificially damping Neptune's eccentricity indirectly damps Uranus' eccentricity due to their secular coupling. {The extent to which Neptune's eccentricity damping affects Uranus depends on the damping timescale and change in eccentricity.} At late stages of evolution, coupling between the latter two planets is important to consider since the planetesimal disk will have dissipated significantly.

\begin{acknowledgments}
{We thank the anonymous reviewer for their comments and suggestions, which led to the improvement of this paper.} We gratefully acknowledge funding from NASA Emerging Worlds grant 80NSSC21K0376. AHR thanks the LSST-DA Data Science Fellowship Program, which is funded by LSST-DA, the Brinson Foundation, and the Moore Foundation; her participation in the program has benefited this work. This material is based upon work supported by the National Science Foundation Graduate Research Fellowship Program under Grant No. 1842400. Any opinions, findings, and conclusions or recommendations expressed in this material are those of the author(s) and do not necessarily reflect the views of the National Science Foundation. KV acknowledges additional funding from NASA grants 80NSSC19K0785 and 80NSSC23K0886. RP acknowledges additional funding from NASA Solar System Observations grant 80NSSC21K0289. We acknowledge use of the lux supercomputer at UC Santa Cruz, funded by NSF MRI grant AST 1828315.  
\end{acknowledgments}

\appendix\label{appendix}

\section{Simulation Parameters}

{The initial conditions for the simulations shown in this paper are provided in Table \ref{all-sim-params}. }

\label{app:sim_params}
\begin{table}
\centering
\begin{tabular}{|c|c|cccc|cccc|cccc|cccc|cccc|c|}
\hline
 & \multicolumn{21}{c}{simulation initial parameters and imposed evolution parameters}&\\
 \cline{2-23}
 & \multicolumn{21}{c}{units: au, Myr, deg }&\\
\hline
Fig. & \multirow{2}{*}{$m$}  & \multicolumn{4}{c|}{$a$ } & \multicolumn{4}{c|}{$e$}& \multicolumn{4}{c|}{$i$ }& \multicolumn{4}{c|}{$\omega$ }&\multicolumn{4}{c|}{$\Omega$} &$f$\\
\cline{3-23}
&  & $a_0$ &  $\Delta a$ &  $\tau_a$ & Eq &$e_0$ &$\Delta e$ & $\tau_e$ & Eq & $i_0$ & $\Delta i$ & $\tau_i$ & Eq & $\omega_0$& $\Delta \omega$ & $\tau_\omega$& Eq & $\Omega_0$ &  $\Delta \Omega$ & $\tau_\Omega$ &  Eq & $f_0$ \\ 
\hline
\ref{fig:1planetex} &$M_J$&5.2&1.8&10&\ref{aeifuncs}a&0.2&-0.1&5&\ref{aeifuncs}c&10&5&4&\ref{aeifuncs}b&50&35&80&\ref{aeifuncs}d&30&60&20&\ref{aeifuncs}b&240\\
\hline
\multirow{2}{*}{\ref{fig:2planetex}} &$M_J$&6&-1&10&\ref{aeifuncs}c&0.2&-0.1&5&\ref{aeifuncs}c&5&-2&20&\ref{aeifuncs}c&50&-&-&-&30&-&-&-&240\\
&$M_N$&23&7&10&\ref{aeifuncs}c&0.1&0.2&5&\ref{aeifuncs}c&10&-8&20&\ref{aeifuncs}c&200&-&-&-&280&-&-&-&250\\
\hline
\multirow{1}{*}{\ref{fig:2planet-incinations}} &$M_J$&6&-&-&-&0.2&-&-&-&10&5&20&\ref{aeifuncs}b&50&-&-&-&30&-&-&-&240\\
a&$M_N$&23&-&-&-&0.1&-&-&-&20&-15&20&\ref{aeifuncs}b&260&-&-&-&287&-&-&-&277\\
\cline{2-23}
\multirow{2}{*}{b}&$M_J$&6&-&-&-&0.2&-&-&-&10&5&5&\ref{aeifuncs}b&50&-&-&-&30&-&-&-&240\\
&$M_N$&23&-&-&-&0.1&-&-&-&20&-15&5&\ref{aeifuncs}b&260&-&-&-&287&-&-&-&277\\
\cline{2-23}
\multirow{2}{*}{c}&$M_J$&6&-&-&-&0.2&-&-&-&10&2&20&\ref{aeifuncs}b&50&-&-&-&30&-&-&-&240\\
&$M_N$&23&-&-&-&0.1&-&-&-&20&-5&20&\ref{aeifuncs}b&260&-&-&-&287&-&-&-&277\\
\hline
\multirow{1}{*}{\ref{fig:twoplanets-wOm-linear}} &$M_J$&6&-&-&-&0.05&-&-&-&1&-&-&-&50&100&50&\ref{aeifuncs}d&30&-&-&-&240\\
\multirow{1}{*}{o} &$M_N$&23&-&-&-&0.05&-&-&-&1&-&-&-&200&-&-&-&70&-&-&-&270\\
\cline{2-23}
\multirow{2}{*}{b}&$M_J$&6&-&-&-&0.05&-&-&-&1&-&-&-&50&-&-&-&30&-&-&-&240\\
&$M_N$&23&-&-&-&0.05&-&-&-&1&-&-&-&200&-&-&-&70&-&-&-&270\\
\hline
\ref{fig:tsiganisearlymig}&$M_J$&5.15&-&-&-&0.052&-&-&-&2&-&-&-&164&-&-&-&163&-&-&-&285\\
\multirow{2}{*}{} &$M_S$&8.75&-&-&-&0.1&-&-&-&4&-&-&-&308&-&-&-&247&-&-&-&73\\
&$M_U$&18.6&0.7&7.2&\ref{aeifuncs}c&0.06&-0.042&1.5&\ref{aeifuncs}c&6&-&-&-&179&-&-&-&276&-&-&-&243\\
&$M_N$&23.7&1.7&11&\ref{aeifuncs}c&0.06&-0.04&1.5&\ref{aeifuncs}c&5&-&-&-&126&-&-&-&10&-&-&-&90\\
\cline{2-23}
\multirow{4}{*}{} &$M_J$&5.15&-&-&-&0.052&-&-&-&2&-&-&-&187&-&-&-&151&-&-&-&36\\
&$M_S$&8.75&-&-&-&0.1&-&-&-&4&-&-&-&89&-&-&-&264&-&-&-&120\\
&$M_U$&19.3&0.4&17&\ref{aeifuncs}c&0.054&-0.024&20&\ref{aeifuncs}c&6&-&-&-&220&-&-&-&308&-&-&-&133\\
&$M_N$&25.5&0.93&9.6&\ref{aeifuncs}c&0.055&-0.033&6&\ref{aeifuncs}c&5&-&-&-&187&-&-&-&36&-&-&-&269\\
\hline
\multirow{4}{*}{\ref{fig:uranusdampmore}} &$M_J$&5.15&-&-&-&0.052&-&-&-&2&-&-&-&122&-&-&-&286&-&-&-&176\\
&$M_S$&8.75&-&-&-&0.1&-&-&-&4&-&-&-&167&-&-&-&160&-&-&-&37\\
&$M_U$&19.3&-&-&-&0.054&-&-&-&6&-&-&-&348&-&-&-&164&-&-&-&278\\
&$M_N$&25.5&0.93&9.6&\ref{aeifuncs}c&0.055&-0.033&6&\ref{aeifuncs}c&5&-&-&-&300&-&-&-&76&-&-&-&49\\
\hline
\multirow{4}{*}{\ref{fig:highUN_edamp}} &$M_J$&5.15&-&-&-&0.052&-&-&-&2&-&-&-&260&-&-&-&65&-&-&-&172\\
&$M_S$&8.75&-&-&-&0.1&-&-&-&4&-&-&-&120&-&-&-&359&-&-&-&28\\
&$M_U$&19.34&-&-&-&0.06&-&-&-&6&-&-&-&136&-&-&-&322&-&-&-&153\\
&$M_N$&25.58&0.9&11.9&\ref{aeifuncs}c&0.065&-0.042&5.1&\ref{aeifuncs}c&5&-&-&-&40&-&-&-&302&-&-&-&236\\
\hline
\multirow{4}{*}{\ref{fig:uranus0.1e}} &$M_J$&5.15&-&-&-&0.052&-&-&-&1&-&-&-&73&-&-&-&260&-&-&-&161\\
&$M_S$&8.75&-&-&-&0.1&-&-&-&2&-&-&-&284&-&-&-&319&-&-&-&286\\
&$M_U$&19.34&-&-&-&0.1&-0.06&5.1&\ref{aeifuncs}c&1&-&-&-&185&-&-&-&191&-&-&-&268\\
&$M_N$&25.58&0.9&11.9&\ref{aeifuncs}c&0.065&-0.042&5.1&\ref{aeifuncs}c&1&-&-&-&177&-&-&-&86&-&-&-&356\\
\hline
\end{tabular}
\caption{{Planet parameters—mass ($m$), semimajor axis ($a$), eccentricity ($e$), inclination ($i$), argument of pericenter ($\omega$), longitude of ascending node ($\Omega$), and true anomoly ($f$)—for simulations illustrated in each figure. If an imposed evolution was used for a planet's orbital element, the columns, $\Delta$, $\tau$, and functional form (Eq \ref{aeifuncs}) are shown. A ``-" denotes the absence of an imposed evolution. 
 The angles $\omega$, $\Omega$, and $f$ for simulations in Figures \ref{fig:tsiganisearlymig}-\ref{fig:uranus0.1e} are produced by a random number generator and are rounded to a whole number here. Because of the chaotic nature of this configuration, i.e. the planets are within the range of another planet's gravitational influence, several simulations with randomized angles are necessary to have at least one evolution similar to these, as discussed in Section \ref{sec:application}. The planet masses $M_J$, $M_S$, $M_U$, $M_N$ are Jupiter, Saturn, Uranus, and Neptune masses, respectively. Semimajor axis, angles, and timescales are in units of au, degree, and Myr, respectively.}}
\label{all-sim-params}
\end{table}

\section{Modifications to the equations of motion}
\label{app:mods}

We modify two N-body integration codes, \texttt{Mercury 6.2} and \texttt{REBOUND}, and confirm that we reproduce the same planetary evolution when applying the same orbital evolution functional forms to identical planetary initial conditions.  

Modifications to \texttt{Mercury 6.2} in the way we are using them, were done by \citet{dawson2012} and \citet{Wolff2012} and are described in the Appendix of \citet{Wolff2012}. We reiterate those modifications in this section with specifics and discuss additional modifications to the equations of motions to allow for artificial evolution of the phase angles: argument of pericenter and longitude of ascending node.
Modifications to \texttt{REBOUND} are explained below as well. We note that these modifications are non-symplectic, even for a symplectic integrator, thus, the modifications must represent small perturbations on the planet for this to be valid. These modifications were tested for the \texttt{MVS} and \texttt{HYBRID} integrators within \texttt{Mercury} and for the \texttt{WHFAST} and \texttt{MERCURIUS} integrators within \texttt{REBOUND};  the tests were run for systems with a star and 1, 2, or 4 planets. 

The modifications in \texttt{REBOUND} are made as follows:
\begin{itemize}
    \item In \texttt{rebound.h}, we created six new members for the structure \texttt{reb\_particle}: \texttt{vusrx}, \texttt{vusry}, \texttt{vusrz}, \texttt{ausrx}, \texttt{ausry}, \texttt{ausrz}. As is default for the other members in the structure, these user velocity and acceleration terms are initialized to nan or zero in \texttt{particle.c} and \texttt{tools.c}.
    
    \item As is default in \texttt{REBOUND}, to turn on extra forces in the simulation, \texttt{sim->additionalforces} must be set to the name of a user-defined function that defines the additional forces on the desired particles. In our case, this function is setting both additional forces \textit{and} velocities, since this is where we add the user velocity and acceleration terms mentioned in bullet one, defined by Equations (\ref{eq:vauser}). For example, in \texttt{sim->additionalforces=migrationforce}, inside the function \texttt{migrationforce}, the user velocity for a given particle, referenced by \texttt{p}, is set by: \texttt{p->vusrx += dxda*adot + dxde*edot + dxdi*idot + dxdom*omdot + dxdOm*Omdot}.

    \item The integration of the motion of particles is typically split up into a few components depending on the type of interactions one is interested in modeling. The update goes in the interaction step (\texttt{Mercurius} goes through \textsc{interaction step}, \textsc{jump step},\textsc{kepler step},\textsc{jump step}, and one final \textsc{interaction step} if the coordinates are not synchronized. Each individual step updates the position and velocity of each particle.) By default, the function \texttt{reb\_integrator\_mercurius\_interaction\_step} in \texttt{integrator\_mercurius.c}, only updates the velocity of the particle with the acceleration. When the \texttt{additionalforces} attribute is turned on, the acceleration includes the additional force. 
    In our case, the user defined force is its own variable: \texttt{ausrx}, \texttt{ausry}, \texttt{ausrz}.
    We modified the function \texttt{reb\_integrator\_mercurius\_interaction\_step} to update the position and velocity of the particle using the additional velocity and acceleration. Specifically, the velocity of the particle is updated with \texttt{ax} as is the default. Then we transform the coordinates from democratic heliocentric coordinates to intertial using \texttt{reb\_integrator\_mercurius\_dh\_to\_inertial(r)} to update the position and velocities using the user defined variables like: 
    
    \texttt{particles[i].vx += dt*particles[i].ausrx
		 \indent particles[i].x  += dt*particles[i].vusrx}.
    
    \noindent Then, we transform the coordinates back to democratic heliocentric with\\ \texttt{reb\_integrator\_mercurius\_inertial\_to\_dh(r)}
    
    \item Adding the six new members to the structure \texttt{REB\_PAR
TICLE}–\texttt{VUSRX}, \texttt{VUSRY}, \texttt{VUSRZ}, \texttt{AUSRX}, \texttt{AUSRY}, \texttt{AUSRZ}–
caused the simulation archive to save simulation data
incorrectly(e.g.,particle parameters were assigned to the
wrong \texttt{REB\_PARTICLE} member). To work around this,
we save simulation outputs into txt or hdf5 files directly. Future modifications will provide a method to save the simulation with the simulation archive.
    
\end{itemize}

The modifications in \texttt{Mercury6.2} are made as follows:
\begin{itemize}
    \item As is default in \texttt{Mercury6.2}, the velocity of an object is updated in the interaction step of a given integrator algorithm. In \texttt{MDT\_HY.FOR}, we modify the interaction step portions (there are two of them) to update the position of the particle as well with the user-defined velocity like \texttt{x(1,j) = x(1,j)  +  hby2 * vusr(1,j)} where \texttt{hby2} is half a time-step. An equivalent line of code is written for the $y$ and $z$ coordinates. In this same routine, we initialize the \texttt{vusr} vector at the beginning. 
    
    \item The \texttt{MFO\_USER.FOR} routine is modified to return a user-defined velocity in addition to the already supported, user-defined acceleration. This routine calculates the user-defined velocity and acceleration terms from Equation \ref{eq:velacc} and applies them to the planets the user desires. Here, we initialize the $\Delta$ values that are part of the $\dot{a}$, $\dot{e}$, etc and so on. 
\end{itemize}

\subsection{Example of the Impact of a User-defined Velocity}
\label{app:accvel}
\begin{figure}
    \centering
    \includegraphics[width=\linewidth]{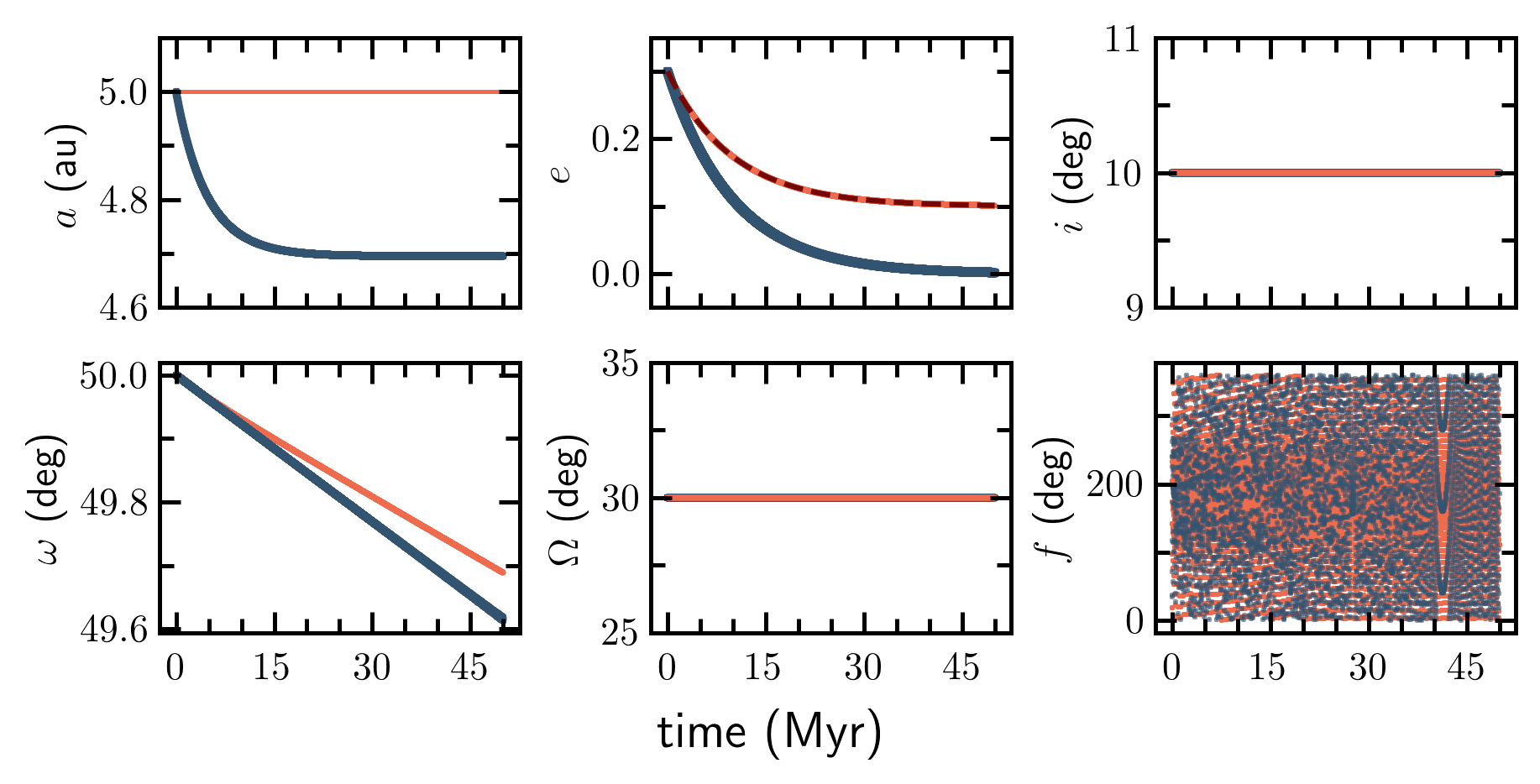}
    \caption{{Two examples of one planet's orbital evolution when the eccentricity is forced to damp exponentially (dashed maroon line). When both the user-defined additional velocities and accelerations are used (orange), the planet's orbital elements evolve independently from each other and only the eccentricity damps as prescribed. When only the user-defined accelerations are used (blue), the planet's semi-major axis also damps, and the eccentricity evolves lower than prescribed.}}
    \label{fig:oneplanet-novusr-yesvusr}
\end{figure}

{Here we discuss how including both additional velocities and accelerations instead of just additional accelerations allows tighter control of the planets' evolution. With both additional velocities and accelerations, the position and velocity of the planet are updated every timestep. Since a set of position and velocity vectors defines a set of six osculating orbital elements, the evolution is done in a consistent way and the elements can be evolved independently from each other. If we only update the 3 components of the velocity vector every timestep (i.e only additional acceleration present), then this does not map consistently to a change in 6 orbital elements. 
To build intuition for this, we provide an example of one planet orbiting a star.  
Figure \ref{fig:oneplanet-novusr-yesvusr} shows the orbital evolution of one planet whose eccentricity is forced to change (dashed line top middle panel) for two cases: (1) both additional velocities and additional accelerations are used (orange) and (2) only additional accelerations (blue). The planet's eccentricity is exponentially damped (Equation \ref{dotaeifuncs}c) with $\tau_e=1$ Myr and $\Delta e= 0.2$. For case (1), the planet's eccentricity evolves exactly as prescribed and the other orbital elements are not disturbed. For case (2), the planet's eccentricity damps further than prescribed, and the semi-major axis also damps. This happens because the additional acceleration applied to change the eccentricity also introduces a torque that changes the energy and hence the semi-major axis.
}

\subsection{Updating Position and Velocity in \texttt{REBOUND}}
\label{app:orb_elem_int}
{The osculating orbital elements that are used for the user velocity and acceleration are calculated once each time the acceleration is calculated (or updated). The order of operations in the \texttt{mercurius} algorithm depends on various integrator-dependent factors (see \citealt{Rein2019}), but in a typical path, the acceleration is calculated twice in a timestep. 
This integrator generally follows a \textit{kick-drift-kick} algorithm, where the momentum is updated for half the timestep, the position is updated for the full timestep, and finally the momentum is updated for the remaining half timestep \citep[see][for a thorough discussion]{Rein2019}. The kick or ``interaction" step in \texttt{mercurius}, integrates the planets' motion due to interactions between the planets, whereas the drift or ``Kepler" step integrates the dominant Keplerian motion around the star. In the publicly available REBOUND code version 4, the interaction step only updates the velocity of the particles given their acceleration, which includes the user-defined acceleration. We modified this function to also update the position of the particles given the user-defined velocities; therefore, the particles experience an extra push in the kick step. This modification is non-symplectic within a symplectic integrator, which is allowed as long as the additional perturbations are small compared to the gravitational effects of the star and other planets. We note that our implemented procedure always updates position and velocity at the same time based on the same set of orbital elements. We tested our modifications for integrators \texttt{WHFAST} and \texttt{MERCURIUS}, thus position and velocity are updated simultaneously in time. If one is interested in following these modifications for other integrators, one should be aware if this is appropriate for that integrator. }

\subsection{Partial Derivatives in User-defined Acceleration and Velocity}
\label{app:derivatrives}
{Here we present the partial derivatives that are in the user-defined acceleration and velocity (Equation \ref{eq:vauser}), which are the additional terms used to artificially evolve a planet's orbital elements. The partial derivatives for the cartesian coordinates ($x$, $y$, $z$, $\dot{x}$, $\dot{y}$, $\dot{z}$) are with respect to the orbital elements, since each orbital element is independent from each other and is time-evolving (i.e. $\dot{a}$, $\dot{e}$, $\dot{I}$, $\dot{\omega}$, $\dot{\Omega}$, $\dot{f}$ may be non-zero). The orbital elements in these equations are derived each timestep from the positions and velocities using standard formulas for osculating orbital elements. The variable $r$—the distance of the planet from the central body—is itself a function of orbital elements, $r = \frac{a(1-e^2)}{1+e\cos{f}}$, so we show partial derivatives for $r$ and $\dot{r}$ as well.}

\begin{equation} \frac{\partial r}{\partial a} = \frac{1-e^2}{1+e\cos{f}} = \frac{r}{a}\end{equation}

\begin{equation} \frac{\partial r}{\partial e} = -\frac{2 a e}{1 + e\cos{f}} - \frac{a(1-e^2)\cos{f}}{(1+e\cos{f})^2} \end{equation}

\begin{equation} \frac{\partial r}{\partial f} = \frac{a(1-e^2)e\sin{f}}{(1+e\cos{f})^2}\end{equation}

\begin{equation} \frac{\partial r\dot{f}}{\partial a} = \frac{-r\dot{f}}{2a} = -\frac{1}{2}\sqrt{\frac{M}{a^3}}\frac{1+e\cos{f}}{\sqrt{1-e^2}}
\end{equation}

\begin{equation} \frac{\partial r\dot{f}}{\partial e} = \frac{r\dot{f}(e+\cos{f})}{(1-e^2)(1+e\cos{f})} = \sqrt{\frac{M}{a}}\frac{e+e\cos{f}}{(1-e^2)^{3/2}}
\end{equation}

\begin{equation} \frac{\partial r\dot{f}}{\partial f} = \frac{-a\sqrt{M/a^3}e\sin{f}}{\sqrt{1-e^2}}\end{equation}

\begin{equation} \frac{\partial \dot{f}}{\partial a} = -3/2 \frac{\sqrt{M/a^5}(1+e\cos{f})^2}{(1-e^2)^{3/2}}\end{equation}

\begin{equation} 
\frac{\partial \dot{f}}{\partial e} = \sqrt{\frac{M}{a^3}} \Big(\frac{2(1+e\cos{f})\cos{f}}{(1-e^2)^{3/2}} + \frac{3e(1+e\cos{f})^2}{(1-e^2)^{5/2}}\Big)
\end{equation}

\begin{equation} \frac{\partial \dot{f}}{\partial f} = - \frac{2\sqrt{M/a^3}(1+e\cos{f})e\sin{f}}{(1-e^2)^{3/2}}\end{equation}

\begin{align}
\frac{\partial \dot{r}}{\partial a} &= \Big(\frac{-2e}{1+e\cos{f}} - \frac{(1-e^2)\cos{f}}{(1+e\cos{f})^2}\Big)\dot{e}\\
& + \Big(\frac{(1-e^2)e\sin{f}}{(1+e\cos{f})^2}\Big)\dot{f} +\frac{a(1-e^2)e\sin{f}}{(1+e\cos{f})^2} \frac{\partial \dot{f}}{\partial a}\notag
\end{align}

\begin{align}
\frac{\partial \dot{r}}{\partial e} &=  \Big(\frac{-2e}{1+e\cos{f}} - \frac{(1-e^2)\cos{f}}{(1+e\cos{f})^2}\Big)\dot{a}\\ 
&+ \Big(\frac{-2a}{(1+e\cos{f})} + \frac{4ae\cos{f}}{(1+e\cos{f})^2} +\frac{2a(1-e^2)(\cos{f})^2}{(1+e\cos{f})^3}\Big)\dot{e}\notag\\ 
&+ \Big(\frac{a(1-3e^2)\sin{f}}{(1+e\cos{f})^2} - \frac{2a(e-e^3)\sin{f}\cos{f}}{(1+e\cos{f})^3}\Big)\dot{f}\notag \\
&+ \Big(\frac{a(1-e^2)e\sin{f}}{(1+e\cos{f})^2}\Big)\frac{\partial \dot{f}}{\partial e}\notag
\end{align}

\begin{align} 
\frac{\partial \dot{r}}{\partial f} &= \frac{(1-e^2)e\sin{f}}{(1+e\cos{f})^2}\dot{a} \\
&+ \Big(\frac{-2ae^2\sin{f}}{(1+e\cos{f})^2} + \frac{a(1-e^2)\sin{f}}{(1+e\cos{f})^2} - \frac{-2a(1-e^2)\cos{f}e\sin{f}}{(1+e\cos{f})^3}\Big) \dot{e}\notag \\
&+ \Big(\frac{a(1-e^2)e\cos{f}}{(1+e\cos{f})^2} + \frac{2a(1-e^2)e^2\sin{f}^2}{(1+e\cos{f})^3}\Big)\dot{f}\notag \\
&+ \Big(\frac{a(1-e^2)e\sin{f}}{(1+e\cos{f})^2} \Big)\frac{\partial \dot{f}}{\partial f}
\end{align}

\begin{equation} \frac{\partial x}{\partial a} = \frac{\partial r}{\partial a}\cos{\Omega}\cos{(\omega + f)} - \frac{\partial r}{\partial a}\cos{i}\sin{\Omega}\sin{(\omega+f)}\end{equation}

\begin{equation} \frac{\partial x}{\partial e} = \frac{\partial r}{\partial e}\cos{\Omega}\cos{(\omega + f)} - \frac{\partial r}{\partial e}\cos{i}\sin{\Omega}\sin{(\omega+f)}\end{equation}

\begin{equation} \frac{\partial x}{\partial i} = r\sin{i}\sin{\Omega}\sin{(\omega + f)}\end{equation}

\begin{equation} \frac{\partial x}{\partial \Omega} = -r\sin{\Omega}\cos{(\omega + f)} - r\cos{i}\cos{\Omega}\sin{(\omega+f)} \end{equation}

\begin{equation} \frac{\partial x}{\partial \omega} = -r\cos{\Omega}\sin{(\omega+f)} - r\cos{i}\sin{\Omega}\cos{(\omega+f)}\end{equation}

\begin{align} \frac{\partial x}{\partial f} &= \frac{\partial r}{\partial f}\cos{\Omega}\cos{(\omega+f)} -r\cos{\Omega}\sin{(\omega+f)} \\
&-\frac{\partial r}{\partial f}\cos{i}\sin{\Omega}\sin{(\omega+f)} - r\cos{i}\sin{\Omega}\cos{(\omega+f)}\notag
\end{align}

\begin{equation} \frac{\partial y}{\partial a} = \frac{\partial r}{\partial a}\sin{\Omega}\cos{(\omega + f)} + \frac{\partial r}{\partial a}\cos{i}\cos{\Omega}\sin{(\omega+f)}\end{equation}

\begin{equation} \frac{\partial y}{\partial e} = \frac{\partial r}{\partial e}\sin{\Omega}\cos{(\omega + f)} + \frac{\partial r}{\partial e}\cos{i}\cos{\Omega}\sin{(\omega+f)}\end{equation}

\begin{equation} \frac{\partial y}{\partial i} = -r\sin{i}\cos{\Omega}\sin{(\omega + f)}\end{equation}

\begin{equation} \frac{\partial y}{\partial \Omega} = r\cos{\Omega}\cos{(\omega + f)} - r\cos{i}\sin{\Omega}\sin{(\omega+f)}\end{equation}

\begin{equation} \frac{\partial y}{\partial \omega} = -r\sin{\Omega}\sin{(\omega+f)} + r\cos{i}\cos{\Omega}\cos{(\omega+f)}\end{equation}

\begin{align} \frac{\partial y}{\partial f} &= \frac{\partial r}{\partial f}\sin{\Omega}\cos{(\omega+f)} -r\sin{\Omega}\sin{(\omega+f)} \\
&+\frac{\partial r}{\partial f}\cos{i}\cos{\Omega}\sin{(\omega+f)} +r\cos{i}\cos{\Omega}\cos{(\omega+f)}\notag
\end{align}

\begin{equation} \frac{\partial z}{\partial a} = \frac{\partial r}{\partial a}\sin{i}\sin{(\omega + f)} \end{equation}

\begin{equation} \frac{\partial z}{\partial e} = \frac{\partial r}{\partial e}\sin{i}\sin{(\omega + f)} \end{equation}

\begin{equation} \frac{\partial z}{\partial i} = r\cos{i}\sin{(\omega + f)} \end{equation}

\begin{equation} \frac{\partial z}{\partial \Omega} = 0.0\end{equation}

\begin{equation} \frac{\partial z}{\partial \omega} = r\sin{i}\cos{(\omega+f)}\end{equation}

\begin{equation} \frac{\partial z}{\partial f} = \frac{\partial r}{\partial f}\sin{i}\sin{(\omega+f)} + r\sin{i}\cos{(\omega+f)}\end{equation}

\begin{align}
\frac{\partial \dot{x}}{\partial a} &= \frac{\partial \dot{r}}{\partial a}\cos{(\Omega)}\cos{(\omega+f)} -\frac{\partial r}{\partial a}\sin{\Omega}\dot{\Omega}\cos{(\omega+f)}\\
&-(\frac{\partial r}{\partial a}\dot{\omega}+\frac{\partial r \dot{f}}{\partial a})\cos{\Omega}\sin{(\omega+f)}-\frac{\partial \dot{r}}{\partial a}\cos{i}\sin{\Omega}\sin{(\omega+f)}\notag\\
&+\frac{\partial r}{\partial a}\sin{i}\dot{I}\sin{\Omega}\sin{(\omega+f)}-\frac{\partial r}{\partial a}\cos{i}\cos{\Omega}\dot{\Omega}\sin{(\omega+f)}\notag\\
&-(\frac{\partial r}{\partial a}\dot{\omega}+\frac{\partial r \dot{f}}{\partial a})\cos{i}\sin{\Omega}\cos{(\omega+f)}\notag
\end{align}

\begin{align}
\frac{\partial \dot{x}}{\partial e}  &= \frac{\partial \dot{r}}{\partial e}\cos{\Omega}\cos{(\omega+f)}-\frac{\partial r}{\partial e}\sin{\Omega}\dot{\Omega}\cos{(\omega+f)}\\
&-(\frac{\partial r}{\partial e}\dot{\omega} + \frac{\partial r\dot{f}}{\partial e})\cos{\Omega}\sin{(\omega+f)}-\frac{\partial \dot{r}}{\partial e}\cos{i}\sin{\Omega}\sin{(\omega+f)}\notag\\
&+\frac{\partial r}{\partial e}\sin{i}\dot{I}\sin{\Omega}\sin{(\omega+f)}-\frac{\partial r}{\partial e}\cos{i}\cos{\Omega}\dot{\Omega}\sin{(\omega+f)}\notag\\
&- (\frac{\partial r}{\partial e}\dot{\omega}+ \frac{\partial r\dot{f}}{\partial e})\cos{i}\sin{\Omega}\cos{(\omega+f)}\notag
\end{align}

\begin{align} 
\frac{\partial \dot{x}}{\partial i} &= \dot{r}\sin{i}\sin{\Omega}\sin{(\omega+f)}+r\cos{i}\dot{I}\sin{\Omega}\sin{(\omega+f)} \\
&+r\sin{i}\cos{\Omega}\dot{\Omega}\sin{(\omega+f)}+ (r\dot{\omega}+r\dot{f})\sin{i}\sin{\Omega}\cos{(\omega+f)}\notag
\end{align}

\begin{align} 
\frac{\partial \dot{x}}{\partial \Omega} &= 
-\dot{r}\sin{\Omega}\cos{(\omega+f)}-r\cos{\Omega}\dot{\Omega}\cos{(\omega+f)} \\
&+(r\dot{\omega}+r\dot{f})\sin{\Omega}\sin{(\omega+f)}-\dot{r}\cos{i}\cos{\Omega}\sin{(\omega+f)}\notag\\
&+r\sin{i}\dot{I}\cos{\Omega}\sin{(\omega+f)}+r\cos{i}\sin{\Omega}\dot{\Omega}\sin{(\omega+f)}\notag\\
&- (r\dot{\omega}+r\dot{f})\cos{i}\cos{\Omega}\cos{(\omega+f)}\notag
\end{align}

\begin{align} 
\frac{\partial \dot{x}}{\partial \omega} &= 
-\dot{r}\cos{\Omega}\sin{(\omega+f)}+ r\sin{\Omega}\dot{\Omega}\sin{(\omega+f)}\notag\\
&- (r\dot{\omega}+r\dot{f})\cos{\Omega}\cos{(\omega+f)}- \dot{r}\cos{i}\sin{\Omega}\cos{(\omega+f)}\notag\\
&+ r\sin{i}\dot{I}\sin{\Omega}\cos{(\omega+f)}- r\cos{i}\cos{\Omega}\dot{\Omega}\cos{(\omega+f)}\notag \\
&+ (r\dot{\omega}+r\dot{f})\cos{i}\sin{\Omega}\sin{(\omega+f)}\notag
\end{align}

\begin{align} \frac{\partial \dot{x}}{\partial f} &= -\dot{r}\cos{\Omega}\sin{(\omega+f)}+r\sin{\Omega}\dot{\Omega}\sin{(\omega+f)}\\
& -(r\dot{\omega}+r\dot{f})\cos{\Omega}\cos{(\omega+f)}-\dot{r}\cos{i}\sin{\Omega}\cos{(\omega+f)}\notag\\
&+r\sin{i}\dot{I}\sin{\Omega}\cos{(\omega+f)}-r\cos{i}\cos{\Omega}\dot{\Omega}\cos{(\omega+f)}\notag \\
&+(r\dot{\omega}+r\dot{f})\cos{i}\sin{\Omega}\sin{(\omega+f)}+\frac{\partial \dot{r}}{\partial f}\cos{\Omega}\cos{(\omega+f)}\notag \\
&-\frac{\partial r}{\partial f}\sin{\Omega}\dot{\Omega}\cos{(\omega+f)}-(\frac{\partial r}{\partial f}\dot{\omega}+\frac{\partial r\dot{f}}{\partial f})\cos{\Omega}\sin{(\omega+f)}\notag\\
&-\frac{\partial\dot{r}}{\partial f}\cos{i}\sin{\Omega}\sin{(\omega+f)} +\frac{\partial r}{\partial f}\sin{i}\dot{I}\sin{\Omega}\sin{(\omega+f)}\notag\\
& -\frac{\partial r}{\partial f} \cos{i}\cos{\Omega}\dot{\Omega}\sin{(\omega+f)} -(\frac{\partial r}{\partial f}\dot{\omega}+\frac{\partial r\dot{f}}{\partial f})\cos{i}\sin{\Omega}\cos{(\omega+f)} \notag
\end{align}

 \begin{align} \frac{\partial \dot{y}}{\partial a} &= \frac{\partial \dot{r}}{\partial a}\sin{\Omega}\cos{(\omega+f)}+\frac{\partial r}{\partial a}\cos{\Omega}\dot{\Omega}\cos{(\omega+f)} \\
 &-(\frac{\partial r}{\partial a}\dot{\omega}+\frac{\partial r\dot{f}}{\partial a})\sin{\Omega}\sin{(\omega+f)}+\frac{\partial \dot{r}}{\partial a}\cos{i}\cos{\Omega}\sin{(\omega+f)}\notag\\
 &-\frac{\partial r}{\partial a}\sin{i}\dot{I}\cos{\Omega}\sin{(\omega+f)}-\frac{\partial r}{\partial a}\cos{i}\sin{\Omega}\dot{\Omega}\sin{(\omega+f)}\notag\\
 &+(\frac{\partial r}{\partial a}\dot{\omega}+\frac{\partial r\dot{f}}{\partial a})\cos{i}\cos{\Omega}\cos{(\omega+f)}\notag
 \end{align}

\begin{align} \frac{\partial \dot{y}}{\partial e} &=\frac{\partial \dot{r}}{\partial e}\sin{\Omega}\cos{(\omega+f)}+\frac{\partial r}{\partial e}\cos{\Omega}\dot{\Omega}\cos{(\omega+f)} \\
&-(\frac{\partial r}{\partial e}\dot{\omega}+\frac{\partial r\dot{f}}{\partial e})\sin{\Omega}\sin{(\omega+f)}+\frac{\partial \dot{r}}{\partial e}\cos{i}\cos{\Omega}\sin{(\omega+f)}\notag\\
&-\frac{\partial r}{\partial e}\sin{i}\dot{I}\cos{\Omega}\sin{(\omega+f)}-\frac{\partial r}{\partial e}\cos{i}\sin{\Omega}\dot{\Omega}\sin{(\omega+f)}\notag\\
&+(\frac{\partial r}{\partial e}\dot{\omega}+\frac{\partial r\dot{f}}{\partial e})\cos{i}\cos{\Omega}\cos{(\omega+f)}\notag
\end{align}

\begin{align} \frac{\partial \dot{y}}{\partial i} &=-\dot{r}\sin{i}\cos{\Omega}\sin{(\omega+f)}-r\cos{i}\dot{I}\cos{\Omega}\sin{(\omega+f)}\\
&+r\sin{i}\sin{\Omega}\dot{\Omega}\sin{(\omega+f)}-(r\dot{\omega}+r\dot{f})\sin{i}\cos{\Omega}\cos{(\omega+f)}\notag
\end{align}

\begin{align}
\frac{\partial \dot{y}}{\partial \Omega} &= \dot{r}\cos{\Omega}\cos{(\omega+f)}-r\sin{\Omega}\dot{\Omega}\cos{(\omega+f)} \\
&-(r\dot{\omega}+r\dot{f})\cos{\Omega}\sin{(\omega+f)}-\dot{r}\cos{i}\sin{\Omega}\sin{(\omega+f)}\notag \\
&+r\sin{i}\dot{I}\sin{\Omega}\sin{(\omega+f)}-r\cos{i}\cos{\Omega}\dot{\Omega}\sin{(\omega+f)}\notag\\
&-(r\dot{\omega}+r\dot{f})\cos{i}\sin{\Omega}\cos{(\omega+f)}\notag
\end{align}

\begin{align} \frac{\partial \dot{y}}{\partial \omega} &=-\dot{r}\sin{\Omega}\sin{(\omega+f)}-r\cos{\Omega}\dot{\Omega}\sin{(\omega+f)}\\
&-(r\dot{\omega}+r\dot{f})\sin{\Omega}\cos{(\omega+f)} +\dot{r}\cos{i}\cos{\Omega}\cos{(\omega+f)}\notag\\
&-r\sin{i}\dot{I}\cos{\Omega}\cos{(\omega+f)}-r\cos{i}\sin{\Omega}\dot{\Omega}\cos{(\omega+f)}\notag\\
&-(r\dot{\omega}+r\dot{f})\cos{i}\cos{\Omega}\sin{(\omega+f)}\notag
\end{align}

\begin{align} \frac{\partial \dot{y}}{\partial f} &= -\dot{r}\sin{\Omega}\sin{(\omega+f)}-r\cos{\Omega}\dot{\Omega}\sin{(\omega+f)}\\
&-(r\dot{\omega}+r\dot{f})\sin{\Omega}\cos{(\omega+f)} +\dot{r}\cos{i}\cos{\Omega}\cos{(\omega+f)}\notag\\
&-r\sin{i}\dot{I}\cos{\Omega}\cos{(\omega+f)}\notag\\
&-r\cos{i}\sin{\Omega}\dot{\Omega}\cos{(\omega+f)}-(r\dot{\omega}+r\dot{f})\cos{i}\cos{\Omega}\sin{(\omega+f)}\notag \\
&+\frac{\partial \dot{r}}{\partial f}\sin{\Omega}\cos{(\omega+f)}+\frac{\partial r}{\partial f}\cos{\Omega}\dot{\Omega}\cos{(\omega+f)}\notag\\
&-(\frac{\partial r}{\partial f}\dot{\omega}+\frac{\partial r\dot{f}}{\partial f})\sin{\Omega}\sin{(\omega+f)}+\frac{\partial \dot{r}}{\partial f}\cos{i}\cos{\Omega}\sin{(\omega+f)}\notag\\
&-\frac{\partial r}{\partial f}\sin{i}\dot{I}\cos{\Omega}\sin{(\omega+f)}-\frac{\partial r}{\partial f}\cos{i}\sin{\Omega}\dot{\Omega}\sin{(\omega+f)}\notag\\
&+(\frac{\partial r}{\partial f}\dot{\omega}+\frac{\partial r\dot{f}}{\partial f})\cos{i}\cos{\Omega}\cos{(\omega+f)}\notag
\end{align}
 
\begin{align}  
\frac{\partial \dot{z}}{\partial a} 
&= \frac{\partial \dot{r}}{\partial a}\sin{i}\sin{(\omega+f)}+\frac{\partial r}{\partial a}\cos{i}\dot{I}\sin{(\omega+f)}\\
&+(\frac{\partial r}{\partial a}\dot{\omega}+\frac{\partial r\dot{f}}{\partial a})\sin{i}\cos{(\omega+f)}\notag
\end{align}

\begin{align}
\frac{\partial \dot{z}}{\partial e} &= \frac{\partial \dot{r}}{\partial e}\sin{i}\sin{(\omega+f)}+\frac{\partial r}{\partial e}\cos{i}\dot{I}\sin{(\omega+f)}\\
&+(\frac{\partial r}{\partial e}\dot{\omega}+\frac{\partial r\dot{f}}{\partial e})\sin{i}\cos{(\omega+f)}\notag
\end{align}
 
\begin{align} 
\frac{\partial \dot{z}}{\partial i} &= \dot{r}\cos{i}\sin{(\omega+f)}-r\sin{i}\dot{I}\sin{(\omega+f)}\\
&+(r\dot{\omega}+r\dot{f})\cos{i}\cos{(\omega+f)}\notag
\end{align}

\begin{equation}  
\frac{\partial \dot{z}}{\partial \Omega}=0.0
\end{equation}

\begin{align} 
\frac{\partial \dot{z}}{\partial \omega}& =\dot{r}\sin{i}\cos{(\omega+f)}+ r\cos{i}\dot{I}\cos{(\omega+f)} \\
&-(r\dot{\omega}+r\dot{f})\sin{i}\sin{(\omega+f)}\notag
\end{align}

\begin{align}
\frac{\partial \dot{z}}{\partial f} &= \dot{r}\sin{i}\cos{(\omega+f)} + r\cos{i}\dot{I}\cos{(\omega+f)} \\
&-(r\dot{\omega}+r\dot{f})\sin{i}\sin{(\omega+f)}+\frac{\partial \dot{r}}{\partial f}\sin{i}\sin{(\omega+f)}\notag\\
&+\frac{\partial r}{\partial f}\cos{i}\dot{I}\sin{(\omega+f)}+(\frac{\partial r}{\partial f}\dot{\omega}+\frac{\partial r\dot{f}}{\partial f})\sin{i}\cos{(\omega+f)}\notag
\end{align}


%

\vspace{5mm}


\software{REBOUND: \citet{Rein2012}, NUMPY: \citet{numpy}, MATPLOTLIB: \citet{matplotlib}, PANDAS: \citet{pandas},
            ASTROPY: \citet{astropy2013,astropy2018,astropy2022} , SCIPY: \citet{scipy} 
          }




\bibliography{forceplanets}{}

\begin{thebibliography}{}
\expandafter\ifx\csname natexlab\endcsname\relax\def\natexlab#1{#1}\fi
\providecommand{\url}[1]{\href{#1}{#1}}
\providecommand{\dodoi}[1]{doi:~\href{http://doi.org/#1}{\nolinkurl{#1}}}
\providecommand{\doeprint}[1]{\href{http://ascl.net/#1}{\nolinkurl{http://ascl.net/#1}}}
\providecommand{\doarXiv}[1]{\href{https://arxiv.org/abs/#1}{\nolinkurl{https://arxiv.org/abs/#1}}}

\bibitem[{{Ali-Dib} {et~al.}(2021){Ali-Dib}, {Marsset}, {Wong}, \& {Dbouk}}]{AliDib2021}
{Ali-Dib}, M., {Marsset}, M., {Wong}, W.-C., \& {Dbouk}, R. 2021, \aj, 162, 19, \dodoi{10.3847/1538-3881/abf6ca}

\bibitem[{{Astropy Collaboration} {et~al.}(2013){Astropy Collaboration}, {Robitaille}, {Tollerud}, {Greenfield}, {Droettboom}, {Bray}, {Aldcroft}, {Davis}, {Ginsburg}, {Price-Whelan}, {Kerzendorf}, {Conley}, {Crighton}, {Barbary}, {Muna}, {Ferguson}, {Grollier}, {Parikh}, {Nair}, {Unther}, {Deil}, {Woillez}, {Conseil}, {Kramer}, {Turner}, {Singer}, {Fox}, {Weaver}, {Zabalza}, {Edwards}, {Azalee Bostroem}, {Burke}, {Casey}, {Crawford}, {Dencheva}, {Ely}, {Jenness}, {Labrie}, {Lim}, {Pierfederici}, {Pontzen}, {Ptak}, {Refsdal}, {Servillat}, \& {Streicher}}]{astropy2013}
{Astropy Collaboration}, {Robitaille}, T.~P., {Tollerud}, E.~J., {et~al.} 2013, \aap, 558, A33, \dodoi{10.1051/0004-6361/201322068}

\bibitem[{{Astropy Collaboration} {et~al.}(2018){Astropy Collaboration}, {Price-Whelan}, {Sip{\H{o}}cz}, {G{\"u}nther}, {Lim}, {Crawford}, {Conseil}, {Shupe}, {Craig}, {Dencheva}, {Ginsburg}, {VanderPlas}, {Bradley}, {P{\'e}rez-Su{\'a}rez}, {de Val-Borro}, {Aldcroft}, {Cruz}, {Robitaille}, {Tollerud}, {Ardelean}, {Babej}, {Bach}, {Bachetti}, {Bakanov}, {Bamford}, {Barentsen}, {Barmby}, {Baumbach}, {Berry}, {Biscani}, {Boquien}, {Bostroem}, {Bouma}, {Brammer}, {Bray}, {Breytenbach}, {Buddelmeijer}, {Burke}, {Calderone}, {Cano Rodr{\'\i}guez}, {Cara}, {Cardoso}, {Cheedella}, {Copin}, {Corrales}, {Crichton}, {D'Avella}, {Deil}, {Depagne}, {Dietrich}, {Donath}, {Droettboom}, {Earl}, {Erben}, {Fabbro}, {Ferreira}, {Finethy}, {Fox}, {Garrison}, {Gibbons}, {Goldstein}, {Gommers}, {Greco}, {Greenfield}, {Groener}, {Grollier}, {Hagen}, {Hirst}, {Homeier}, {Horton}, {Hosseinzadeh}, {Hu}, {Hunkeler}, {Ivezi{\'c}}, {Jain}, {Jenness}, {Kanarek}, {Kendrew}, {Kern}, {Kerzendorf}, {Khvalko}, {King}, {Kirkby}, {Kulkarni},
  {Kumar}, {Lee}, {Lenz}, {Littlefair}, {Ma}, {Macleod}, {Mastropietro}, {McCully}, {Montagnac}, {Morris}, {Mueller}, {Mumford}, {Muna}, {Murphy}, {Nelson}, {Nguyen}, {Ninan}, {N{\"o}the}, {Ogaz}, {Oh}, {Parejko}, {Parley}, {Pascual}, {Patil}, {Patil}, {Plunkett}, {Prochaska}, {Rastogi}, {Reddy Janga}, {Sabater}, {Sakurikar}, {Seifert}, {Sherbert}, {Sherwood-Taylor}, {Shih}, {Sick}, {Silbiger}, {Singanamalla}, {Singer}, {Sladen}, {Sooley}, {Sornarajah}, {Streicher}, {Teuben}, {Thomas}, {Tremblay}, {Turner}, {Terr{\'o}n}, {van Kerkwijk}, {de la Vega}, {Watkins}, {Weaver}, {Whitmore}, {Woillez}, {Zabalza}, \& {Astropy Contributors}}]{astropy2018}
{Astropy Collaboration}, {Price-Whelan}, A.~M., {Sip{\H{o}}cz}, B.~M., {et~al.} 2018, \aj, 156, 123, \dodoi{10.3847/1538-3881/aabc4f}

\bibitem[{{Astropy Collaboration} {et~al.}(2022){Astropy Collaboration}, {Price-Whelan}, {Lim}, {Earl}, {Starkman}, {Bradley}, {Shupe}, {Patil}, {Corrales}, {Brasseur}, {N{\"o}the}, {Donath}, {Tollerud}, {Morris}, {Ginsburg}, {Vaher}, {Weaver}, {Tocknell}, {Jamieson}, {van Kerkwijk}, {Robitaille}, {Merry}, {Bachetti}, {G{\"u}nther}, {Aldcroft}, {Alvarado-Montes}, {Archibald}, {B{\'o}di}, {Bapat}, {Barentsen}, {Baz{\'a}n}, {Biswas}, {Boquien}, {Burke}, {Cara}, {Cara}, {Conroy}, {Conseil}, {Craig}, {Cross}, {Cruz}, {D'Eugenio}, {Dencheva}, {Devillepoix}, {Dietrich}, {Eigenbrot}, {Erben}, {Ferreira}, {Foreman-Mackey}, {Fox}, {Freij}, {Garg}, {Geda}, {Glattly}, {Gondhalekar}, {Gordon}, {Grant}, {Greenfield}, {Groener}, {Guest}, {Gurovich}, {Handberg}, {Hart}, {Hatfield-Dodds}, {Homeier}, {Hosseinzadeh}, {Jenness}, {Jones}, {Joseph}, {Kalmbach}, {Karamehmetoglu}, {Ka{\l}uszy{\'n}ski}, {Kelley}, {Kern}, {Kerzendorf}, {Koch}, {Kulumani}, {Lee}, {Ly}, {Ma}, {MacBride}, {Maljaars}, {Muna}, {Murphy}, {Norman},
  {O'Steen}, {Oman}, {Pacifici}, {Pascual}, {Pascual-Granado}, {Patil}, {Perren}, {Pickering}, {Rastogi}, {Roulston}, {Ryan}, {Rykoff}, {Sabater}, {Sakurikar}, {Salgado}, {Sanghi}, {Saunders}, {Savchenko}, {Schwardt}, {Seifert-Eckert}, {Shih}, {Jain}, {Shukla}, {Sick}, {Simpson}, {Singanamalla}, {Singer}, {Singhal}, {Sinha}, {Sip{\H{o}}cz}, {Spitler}, {Stansby}, {Streicher}, {{\v{S}}umak}, {Swinbank}, {Taranu}, {Tewary}, {Tremblay}, {de Val-Borro}, {Van Kooten}, {Vasovi{\'c}}, {Verma}, {de Miranda Cardoso}, {Williams}, {Wilson}, {Winkel}, {Wood-Vasey}, {Xue}, {Yoachim}, {Zhang}, {Zonca}, \& {Astropy Project Contributors}}]{astropy2022}
{Astropy Collaboration}, {Price-Whelan}, A.~M., {Lim}, P.~L., {et~al.} 2022, \apj, 935, 167, \dodoi{10.3847/1538-4357/ac7c74}

\bibitem[{{Binney} \& {Tremaine}(2008)}]{Binney2008}
{Binney}, J., \& {Tremaine}, S. 2008, {Galactic Dynamics: Second Edition}

\bibitem[{{Chambers}(1999)}]{Chambers:1999}
{Chambers}, J.~E. 1999, \mnras, 304, 793, \dodoi{10.1046/j.1365-8711.1999.02379.x}

\bibitem[{Chambers(2009)}]{chambers2009}
Chambers, J.~E. 2009, Annual Review of Earth and Planetary Sciences, 37, 321, \dodoi{https://doi.org/10.1146/annurev.earth.031208.100122}

\bibitem[{{Clement} {et~al.}(2018){Clement}, {Kaib}, {Raymond}, \& {Walsh}}]{Clement2018}
{Clement}, M.~S., {Kaib}, N.~A., {Raymond}, S.~N., \& {Walsh}, K.~J. 2018, \icarus, 311, 340, \dodoi{10.1016/j.icarus.2018.04.008}

\bibitem[{Dawson \& Murray-Clay(2012)}]{dawson2012}
Dawson, R.~I., \& Murray-Clay, R. 2012, The Astrophysical Journal, 750, 43, \dodoi{10.1088/0004-637X/750/1/43}

\bibitem[{{de Sousa} {et~al.}(2020){de Sousa}, {Morbidelli}, {Raymond}, {Izidoro}, {Gomes}, \& {Vieira Neto}}]{deSousa:2020}
{de Sousa}, R.~R., {Morbidelli}, A., {Raymond}, S.~N., {et~al.} 2020, Icar, 339, 113605, \dodoi{10.1016/j.icarus.2019.113605}

\bibitem[{{Fernandez} \& {Ip}(1984)}]{Fernandez1984}
{Fernandez}, J.~A., \& {Ip}, W.~H. 1984, \icarus, 58, 109, \dodoi{10.1016/0019-1035(84)90101-5}

\bibitem[{{Ford} \& {Chiang}(2007)}]{Ford2007}
{Ford}, E.~B., \& {Chiang}, E.~I. 2007, \apj, 661, 602, \dodoi{10.1086/513598}

\bibitem[{Gladman \& Volk(2021)}]{Gladman:2021}
Gladman, B., \& Volk, K. 2021, Annual Review of Astronomy and Astrophysics, 59, 203, \dodoi{10.1146/annurev-astro-120920-010005}

\bibitem[{{Goldreich} {et~al.}(2004){Goldreich}, {Lithwick}, \& {Sari}}]{Goldreich2004}
{Goldreich}, P., {Lithwick}, Y., \& {Sari}, R. 2004, \araa, 42, 549, \dodoi{10.1146/annurev.astro.42.053102.134004}

\bibitem[{{Greenberg} {et~al.}(2013){Greenberg}, {Van Laerhoven}, \& {Barnes}}]{Greenberg2013}
{Greenberg}, R., {Van Laerhoven}, C., \& {Barnes}, R. 2013, Celestial Mechanics and Dynamical Astronomy, 117, 331, \dodoi{10.1007/s10569-013-9518-3}

\bibitem[{{Hahn} \& {Malhotra}(2005)}]{Hahn:2005}
{Hahn}, J.~M., \& {Malhotra}, R. 2005, AJ, 130, 2392, \dodoi{10.1086/452638}

\bibitem[{{Hansen} \& {Murray}(2015)}]{Hansen2015}
{Hansen}, B. M.~S., \& {Murray}, N. 2015, \mnras, 448, 1044, \dodoi{10.1093/mnras/stv049}

\bibitem[{Hunter(2007)}]{matplotlib}
Hunter, J.~D. 2007, Computing in Science \& Engineering, 9, 90, \dodoi{10.1109/MCSE.2007.55}

\bibitem[{{Lee} \& {Peale}(2002)}]{Lee2002}
{Lee}, M.~H., \& {Peale}, S.~J. 2002, \apj, 567, 596, \dodoi{10.1086/338504}

\bibitem[{{Levison} {et~al.}(2011){Levison}, {Morbidelli}, {Tsiganis}, {Nesvorn{\'y}}, \& {Gomes}}]{Levison2011}
{Levison}, H.~F., {Morbidelli}, A., {Tsiganis}, K., {Nesvorn{\'y}}, D., \& {Gomes}, R. 2011, \aj, 142, 152, \dodoi{10.1088/0004-6256/142/5/152}

\bibitem[{Levison {et~al.}(2008)Levison, Morbidelli, Van~Laerhoven, Gomes, \& Tsiganis}]{Levison2008}
Levison, H.~F., Morbidelli, A., Van~Laerhoven, C., Gomes, R., \& Tsiganis, K. 2008, Icarus, 196, 258, \dodoi{10.1016/j.icarus.2007.11.035}

\bibitem[{Malhotra(1995)}]{malhotra1995}
Malhotra, R. 1995, The Astronomical Journal, 110, 420, \dodoi{10.1086/117532}

\bibitem[{Morbidelli \& Nesvorný(2020)}]{morbidelli2020}
Morbidelli, A., \& Nesvorný, D. 2020, The Trans-Neptunian Solar System, 25, \dodoi{10.1016/B978-0-12-816490-7.00002-3}

\bibitem[{{Murray} \& {Dermott}(1999)}]{SSDBook}
{Murray}, C.~D., \& {Dermott}, S.~F. 1999, {Solar system dynamics} ({Cambridge University Press})

\bibitem[{{Nesvorn{\'y}}(2015)}]{Nesvorny:2015}
{Nesvorn{\'y}}, D. 2015, \aj, 150, 68, \dodoi{10.1088/0004-6256/150/3/68}

\bibitem[{{Nesvorn{\'y}}(2018)}]{Nesvorny:2018}
---. 2018, ARA\&A, 56, 137, \dodoi{10.1146/annurev-astro-081817-052028}

\bibitem[{{Nesvorn{\'y}} \& {Vokrouhlick{\'y}}(2016)}]{Nesvorny2016}
{Nesvorn{\'y}}, D., \& {Vokrouhlick{\'y}}, D. 2016, \apj, 825, 94, \dodoi{10.3847/0004-637X/825/2/94}

\bibitem[{{Rein} \& {Liu}(2012)}]{Rein2012}
{Rein}, H., \& {Liu}, S.~F. 2012, \aap, 537, A128, \dodoi{10.1051/0004-6361/201118085}

\bibitem[{Rein \& Tamayo(2015)}]{rein2015}
Rein, H., \& Tamayo, D. 2015, Monthly Notices of the Royal Astronomical Society, 452, 376, \dodoi{10.1093/mnras/stv1257}

\bibitem[{{Rein} {et~al.}(2019){Rein}, {Hernandez}, {Tamayo}, {Brown}, {Eckels}, {Holmes}, {Lau}, {Leblanc}, \& {Silburt}}]{Rein2019}
{Rein}, H., {Hernandez}, D.~M., {Tamayo}, D., {et~al.} 2019, \mnras, 485, 5490, \dodoi{10.1093/mnras/stz769}

\bibitem[{{Stewart} \& {Ida}(2000)}]{Stewart2000}
{Stewart}, G.~R., \& {Ida}, S. 2000, \icarus, 143, 28, \dodoi{10.1006/icar.1999.6242}

\bibitem[{{Tamayo} {et~al.}(2020){Tamayo}, {Rein}, {Shi}, \& {Hernandez}}]{Tamayo2020}
{Tamayo}, D., {Rein}, H., {Shi}, P., \& {Hernandez}, D.~M. 2020, \mnras, 491, 2885, \dodoi{10.1093/mnras/stz2870}

\bibitem[{{Tsiganis} {et~al.}(2005){Tsiganis}, {Gomes}, {Morbidelli}, \& {Levison}}]{Tsiganis:2005}
{Tsiganis}, K., {Gomes}, R., {Morbidelli}, A., \& {Levison}, H.~F. 2005, Natur, 435, 459, \dodoi{10.1038/nature03539}

\bibitem[{van~der Walt {et~al.}(2011)van~der Walt, Colbert, \& Varoquaux}]{numpy}
van~der Walt, S., Colbert, S.~C., \& Varoquaux, G. 2011, Computing in Science Engineering, 13, 22, \dodoi{10.1109/MCSE.2011.37}

\bibitem[{Virtanen {et~al.}(2020)Virtanen, Gommers, Oliphant, Haberland, Reddy, Cournapeau, Burovski, Peterson, Weckesser, Bright, {van der Walt}, Brett, Wilson, Millman, Mayorov, Nelson, Jones, Kern, Larson, Carey, Polat, Feng, Moore, {VanderPlas}, Laxalde, Perktold, Cimrman, Henriksen, Quintero, Harris, Archibald, Ribeiro, Pedregosa, {van Mulbregt}, \& {SciPy 1.0 Contributors}}]{scipy}
Virtanen, P., Gommers, R., Oliphant, T.~E., {et~al.} 2020, Nature Methods, 17, 261, \dodoi{10.1038/s41592-019-0686-2}

\bibitem[{{Volk} \& {Malhotra}(2019)}]{volk2019}
{Volk}, K., \& {Malhotra}, R. 2019, AJ, 158, 64, \dodoi{10.3847/1538-3881/ab2639}

\bibitem[{{W}es {M}c{K}inney(2010)}]{pandas}
{W}es {M}c{K}inney. 2010, in {P}roceedings of the 9th {P}ython in {S}cience {C}onference, ed. {S}t\'efan van~der {W}alt \& {J}arrod {M}illman, 56 -- 61, \dodoi{10.25080/Majora-92bf1922-00a}

\bibitem[{{Wisdom} \& {Holman}(1991)}]{Wisdom1991}
{Wisdom}, J., \& {Holman}, M. 1991, \aj, 102, 1528, \dodoi{10.1086/115978}

\bibitem[{{Wolff} {et~al.}(2012){Wolff}, {Dawson}, \& {Murray-Clay}}]{Wolff2012}
{Wolff}, S., {Dawson}, R.~I., \& {Murray-Clay}, R.~A. 2012, \apj, 746, 171, \dodoi{10.1088/0004-637X/746/2/171}

\end{thebibliography}
\bibliographystyle{aasjournal}



\end{document}